\documentclass[12pt,a4paper]{report}

\usepackage[top=0.70in, bottom=0.70in, left=0.8in,right=0.80in]{geometry} % setting the page alignment with this package
\usepackage[pdftex]{graphicx} %for embedding images
\usepackage[final]{pdfpages} %for embedding another pdf, remove if not required
\usepackage{float} %used for figure placement with H as a parameter
\usepackage{pslatex} % for times new roman, old package, but works
\usepackage{array} % for making text bold in table
\usepackage{setspace}
\usepackage{float}
\usepackage{enumerate}
\usepackage{longtable}
\newcommand{\eml}{\textsc{Embedded Matlab}}
\newcommand{\prover}{\textsc{Prover}}

\newcommand{\ccsl}{\textsc{Ccsl}}
\newcommand{\tdl}{\textsc{Tadl}}

\newcommand{\gt}[1]{\texttt{#1}}

\newcommand{\fp}{\textsc{$f_p$}}
\newcommand{\av}{\textsc{AV}}

\newcommand{\uppaal}{\textsc{Uppaal}}

\newcommand{\smc}{\textsc{Uppaal-smc}}

\newcommand{\simu}{\textsc{Simulink}}
\newcommand{\staf}{\textsc{Stateflow}}
\newcommand{\sldv}{\textsc{Simulink  Design Verifier}}
\newcommand{\sdv}{\textsc{SDV}}

\newcommand{\ed}{\textsc{East-adl}}
\newcommand{\auto}{\textsc{AUTOSAR}}

\newcommand{\marte}{\textsc{MARTE}}

\usepackage{makeidx}  % allows for indexgeneration
\usepackage{url}
\usepackage{graphicx}
\usepackage{subfigure}
\usepackage{makecell}
\usepackage{cite}
\usepackage{listings}
\usepackage{paralist}
\let\subparagraph\paragraph
\usepackage{cite}
\usepackage{amsfonts, amsmath, amssymb}
\usepackage{mathptmx}
\usepackage{algorithm}
\usepackage{algorithmic}
\usepackage{epstopdf}
\usepackage{float}
\usepackage{url}
\usepackage{enumerate}
\usepackage{color}
\usepackage{multirow}
\usepackage{verbatim}
\usepackage{bm}
\usepackage{longtable}
\usepackage[figuresright]{rotating}
\usepackage{geometry}
\usepackage{authblk}
\newtheorem{mydef}{Definition}

\usepackage[font=small,labelfont=bf]{caption}

\usepackage{listings}
\usepackage{color}

\definecolor{dkgreen}{rgb}{0,0.6,0}
\definecolor{gray}{rgb}{0.5,0.5,0.5}
\definecolor{mauve}{rgb}{0.58,0,0.82}

\usepackage{fancyhdr}
\fancypagestyle{plain}{%
\fancyfoot[L]{\emph{}} % except the center
\fancyfoot[R]{\thepage}

}

\usepackage{pgf}
\usepackage{pgfpages}

\pgfpagesdeclarelayout{boxed}
{
  \edef\pgfpageoptionborder{0pt}
}
{
  \pgfpagesphysicalpageoptions
  {%
    logical pages=1,%
  }
  \pgfpageslogicalpageoptions{1}
  {
    border code=\pgfsetlinewidth{2pt}\pgfstroke,%
    border shrink=\pgfpageoptionborder,%
    resized width=.95\pgfphysicalwidth,%
    resized height=.95\pgfphysicalheight,%
    center=\pgfpoint{.5\pgfphysicalwidth}{.5\pgfphysicalheight}%
  }%
}
\begin{document}
\renewcommand\bibname{References}
\lhead{ }

%FROM HERE YOUR PAGES START GETTING ADDED

% includes the cover page
%\input{project/cover.tex}
%\newpage

%\input{project/title.tex}
\newpage
\begin{center}
\thispagestyle{empty}
\setlength{\voffset}{2in}
%\LARGE{\textbf{TECHNICAL REPORT}}\\
\vspace{2cm}
\LARGE{\textbf{Formal Specification \& Analysis of Autonomous Systems in PrCCSL/Simulink
Design Verifier}}\\
\vspace{2cm}
{\Large{\textbf{Eun-Young Kang$^{12}$, Li Huang$^{2}$}}}\\
\vspace{0.3cm}
\large{\textbf{$^1$PReCISE Research Centre,
University of Namur, Belgium}}\\
\vspace{0.3cm}
\large{\textbf{$^2$School of Data and Computer Science, \\Sun Yat-sen University, Guangzhou, China}}\\
\vspace{0.3cm}
\Large{{eykang@fundp.ac.be}}\\
\Large{{huangl223@mail2.sysu.edu.cn}}\\
\vspace{3cm}
%All rights reserved. This report may not be reproduced in whole or in part, by photocopying or other means, without the permission of the author.
\newpage

\end{center}
\newpage

% includes the certificate page
%\input{project/certificate.tex}
%\newpage

% includes the acknowledgements page
%\input{project/acknowledgements.tex}
%\newpage

%\input{project/abstract.tex} % adds the Research Methodology page
\begin{center}
\thispagestyle{empty}
\vspace{2cm}
\LARGE{\textbf{ABSTRACT}}\\[1.0cm]
\end{center}
\thispagestyle{empty}
\large{\paragraph{}
Modeling and analysis of timing constraints is crucial in automotive systems. \ed\ is a
domain specific architectural language dedicated to safety-critical
automotive embedded system design.
In most cases, a bounded number of violations of
timing constraints in systems would not lead to system failures when the results of
the violations are negligible, called Weakly-Hard (WH).
We have previously specified \ed\
timing constraints in Clock Constraint Specification Language (\ccsl) and
transformed timed behaviors in \ccsl\ into formal models amenable to model
checking. Previous work is extended in this paper by including support
for probabilistic analysis of timing constraints in the context of WH: Probabilistic extension of \ccsl, called Pr\ccsl, is
defined and the \ed\ timing constraints with stochastic properties
are specified in Pr\ccsl. The semantics of the extended constraints in Pr\ccsl\ is translated
into \emph{Proof Objective Models} that can
be verified using \sldv. Furthermore, a set of mapping rules is proposed to facilitate guarantee of translation.
Our approach is demonstrated
on an autonomous traffic sign recognition vehicle case study.}

\newpage

%TABLE OF CONTENTS AND LIST OF FIGURES ARE AUTOMATICALLY ADDED BY FOLLOWING COMMANDS
%ADD FIGURE OF TABLES IF YOU NEED TO, CHECK DOCUMENTATION
\pagenumbering{roman} %numbering before main content starts

%To reset the Header & Footer for TOC and LOF
\pagestyle{empty}
\addtocontents{toc}{\protect\thispagestyle{empty}}
\tableofcontents % adds Index Page

\addtocontents{lof}{\protect\thispagestyle{empty}}
\listoffigures % adds List of Figures
\cleardoublepage

%And reset back the settings we choose for Header and Footer
\pagestyle{fancy}

\newpage
\pagenumbering{arabic} %reset numbering to normal for the main content

\chapter{Introduction}
Software development for Cyber-Physical Systems (CPS) requires both
functional and non-functional quality assurance to guarantee that CPS
operate in a safety-critical context under timing constraints.
Automotive electric/electronic systems are ideal examples of CPS in
which the software controllers interact with physical environments. The
continuous time behaviors  of those systems often rely on complex
dynamics as well as on stochastic behaviors. Formal verification and
validation (V\&V) technologies are indispensable and highly recommended
for development of safe and reliable automotive systems
\cite{iso26262,iec61508}.

Conventional formal analysis of timing models addresses worst case
designs, typically used for hard deadlines in safety critical systems,
however, there is great incentive to include ``less-than-worst-case''
designs with a view to improving efficiency without affecting the quality
of timing analysis in the systems. The challenge is the definition of
suitable model semantics providing reliable predictions of \emph{system
timing}, given the timing of individual components and their
compositions. While the standard worst case models are well understood
in this respect, the behavior and the expressiveness of
``less-than-worst-case' models is far less investigated. In most cases,
a bounded number of violations of timing constraints in
systems would not lead to system failures when the results of the
violations are negligible, called Weakly-Hard (WH)
\cite{Nicolau1988Specification, Bernat2001Weakly}. In this paper, we
propose a formal probabilistic modeling and analysis technique by
extending the known concept of WH constraints to what is called
``typical'' worst case model and analysis.

\ed\ (Electronics Architecture and Software Technology- Architecture
Description Language) \cite{EAST-ADL,maenad}, aligned with \auto\
(Automotive Open System Architecture) standard \cite{autosar}, is the model-based development approach for the architectural modeling
of safety-critical automotive embedded systems.  A system in \ed\ is
described by {\gt{Functional Architectures (FA)}} at different
abstraction levels. The {\gt{FA}} are composed of a number of
interconnected \emph{functionprototypes} (\fp), and the \fp s have ports
and connectors for communication. \ed\ relies on external tools for the
analysis of specifications related to requirements. For example,
behavioral description in \ed\ is captured in external tools, i.e.,
\simu/\staf \cite{slsf}.  The latest release of \ed\ has adopted the time model
proposed in the Timing Augmented Description Language (\tdl)
\cite{TADL2}. \tdl\ expresses and
composes the basic timing constraints, i.e., repetition rates,
end-to-end delays, and synchronization constraints. The time model
of \tdl\ specializes the time model of \marte, the UML profile for
Modeling and Analysis of Real-Time and Embedded systems
\cite{MARTE}. \marte\ provides \ccsl, a time model and a Clock
Constraint Specification Language, that supports specification of
both logical and dense timing constraints for \marte\ models, as well
as functional causality constraints \cite{ccsl_correctness}.

We have previously specified non-functional properties (timing and
energy constraints) of automotive systems specified in \ed\ and
\marte/\ccsl, and proved the correctness of specification by mapping the
semantics of the constraints into \uppaal\ models for model checking
\cite{ksac14}.
Previous work is extended in this paper by including support
for probabilistic analysis of timing constraints of automotive systems
in the context WH: \begin{inparaenum} \item Probabilistic extension of
\ccsl, called Pr\ccsl, is
defined and the \ed/\tdl\ timing constraints with stochastic properties
are specified in Pr\ccsl; \item The semantics of the extended
constraints in Pr\ccsl\ is translated into verifiable \emph{Proof
Objective Models} (POMs) for formal verification using \sldv\ (SDV)
\cite{SLDV}; \item A set of mapping rules is
proposed to facilitate guarantee of translation. \end{inparaenum} Our
approach is demonstrated on an autonomous traffic sign recognition
vehicle (AV) case study.

The paper is organized as follows: Sec. \ref{sec:preli} presents an overview
of \ccsl, \simu/\staf\ and SDV. The AV is introduced
as a running example in Sec. \ref{sec:case-study}. Sec. \ref{sec:prccsl} presents the formal definition of
Pr\ccsl\ and Sec. \ref{sec: specification} demonstrates the specification of \ed\ timing constraints in \ccsl/Pr\ccsl.  Sec. \ref{sec:model} describes a set of translation patterns from
\ccsl/Pr\ccsl\ to POMs and how our approaches provide support for formal
analysis at the design level. The modeling of AV system and its environments in S/S are illustrated in Sec. \ref{sec: av_ss}.
The applicability of our method is demonstrated by performing
verification on the AV case study in Sec. \ref{sec:verification}. Sec. \ref{sec: related work} and  Sec. \ref{sec: conclusion} present
related work and the conclusion.

\chapter{Preliminary}
\label{sec:preli}

In our framework, we consider a subset of \ccsl\ and its extension with stochastic properties that is sufficient to specify \ed\ timing constraints in the context of WH automotive systems. \simu\ and \eml\ (EML) are utilized for modeling purposes, and V\&V are performed by the \simu\ built-in verification tool, \sldv\ (\sdv).

\section{Clock Constraint Specification Language (\ccsl)}
\ccsl\ \cite{ccsl_correctness,andre2009syntax} clocks describe events in a system and measure occurrences of the events. The physical time is represented by a dense clock (with a base) and discretized into a logical clock. $idealClock$ is a predefined dense clock whose unit is \emph{second}. We define a universal clock $ms$ based on $idealClock$: $ms$ = $idealClock$ {\gt{discretizedBy}} 0.001, where $ms$ is a periodic clock that ticks every 1 millisecond. A step is a tick of the universal clock. Hence the length of one step is 1 millisecond in this paper.

\ccsl\ provides two types of clock constraints, \emph{relation} and \emph{expression}: A \emph{relation} limits the occurrences among different events/clocks. Let $C$ be a set of clocks, $c1, c2 \in C$, {\gt{Coincidence}} relation ($c1$ $\equiv$ $c2$) specifies that two clocks must tick simultaneously.  {\gt {Precedence}} relation ($c1 \prec c2$) limits that $c1$ runs faster than $c2$, i.e., $\forall k \in \mathbb{N^{+}}$, where $\mathbb{N^{+}}$ is the set of positive natural numbers, the $k^{th}$ tick of $c1$ must occur prior to the $k^{th}$ tick of $c2$. {\gt{Causality}} relation ($c1 \preceq c2$) represents a relaxed version of {\gt {Precedence}}, allowing the two clocks to tick at the same time. {\gt{Subclock}} ($c1$ $\subseteq$ $c2$) indicates the relation between two clocks, \emph{superclock} ($c1$) and \emph{subclock} ($c2$), s.t. each tick of the subclock must correspond to a tick of its superclock at the same step. {\gt{Exclusion}} ($c1$ \# $c2$) prevents the instants of two clocks from being coincident. An \emph{expression} derives new clocks from the already defined clocks: {\gt{PeriodicOn}} builds a new clock found on a \emph{base} clock and a \emph{period} parameter, s.t., the instants of the new clocks are separated by a number of instants of the \emph{base} clock. The number is given as \emph{period}. {\gt{DelayFor}} results in a clock by delaying the \emph{base} clock for a given number of ticks of a \emph{reference} clock. {\gt{Infimum}}, denoted {\gt{Inf}}, is defined as the slowest clock that is faster than both $c1$ and $c2$.  {\gt{Supremum}}, denoted {\gt{Sup}}, is defined as the fastest clock that is slower than $c1$ and $c2$.

\section{Simulink and SDV}

\begin{figure}[htbp]
\centerline{{\includegraphics[width=5in]{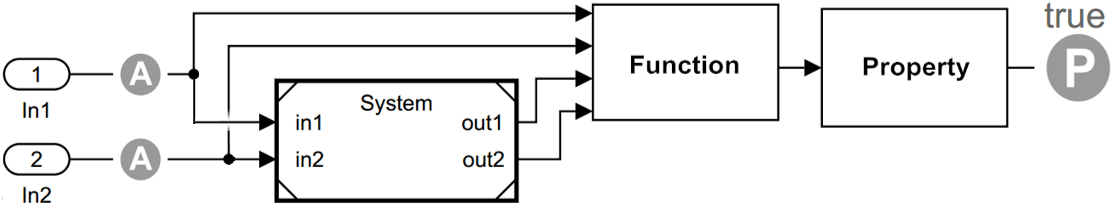}
}}
\caption{General verification models in SDV}
\label{fig_general_verification_model}
\hfil
\end{figure}

\simu\ \cite{slsf} is a synchronous data flow language, which provides different types of \emph{blocks}  for modeling and simulation of dynamic systems and code generation. \simu\ supports the definition of custom blocks via \staf\ diagrams or \emph{user-defined function} blocks written in EML, C, and C++.
SDV is a formal verification tool that performs reachability analysis on \simu/\staf\ (S/S) model with \prover\ plugin. The satisfiability of each reachable state is determined by a SAT solver. A proof objective model is specified in \simu/SDV and illustrated in Fig.\ref{fig_general_verification_model}. A set of data (predicates) on the input flows of \emph{System} is constrained via {\scriptsize $\ll$}Proof Assumption{\scriptsize $\gg$} blocks during proof construction. A set of proof objectives are constructed via a function $F$ block and the output of $F$ is specified as input to a property $P$ block.
$P$ passes its output signal to an {\scriptsize $\ll$}Assertion{\scriptsize $\gg$} block and returns \emph{true} when the predicates set on the input data flows of the outline model are satisfied. Whenever {\scriptsize $\ll$}Assertion{\scriptsize $\gg$} is utilized, SDV verifies whether the specified input data flow is always \emph{true}.
Any failed proof attempt ends in the generation of a counterexample representing an execution path to an invalid state. A harness model is generated to analyze the counterexample and refine the model.

\chapter{Running Example: Traffic Sign Recognition Vehicle}
\label{sec:case-study}
\begin{figure}[t]
  \centering
  \includegraphics[width=5.5in]{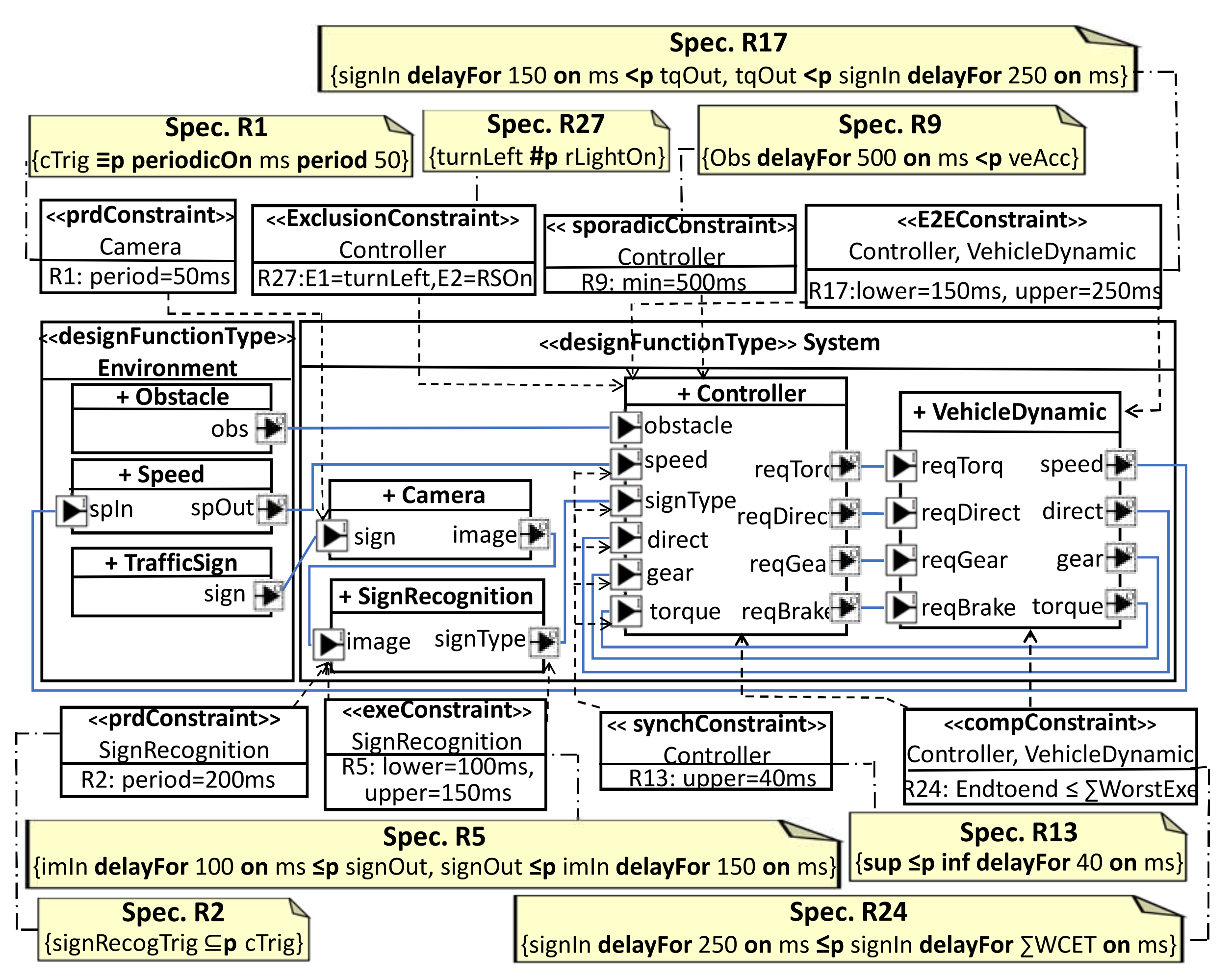}
  \caption{\av\ in \ed\ augmented with \tdl\ timing constraints ({\gt{R.ID}}), specified in Pr\ccsl\ ({\gt{Spec.R.ID)}}}
  \label{fig:East-adl model}
\end{figure}

An autonomous vehicle (\av) \cite{mvv,sscps} application using Traffic Sign Recognition is adopted to illustrate our approach. The \av\ reads the road signs, e.g., ``speed limit'' or ``right/left turn'', and adjusts speed and movement accordingly. The functionality of \av, augmented with timing constraints and viewed as {\gt{Functional Design Architecture (FDA)}} ({\gt{designFunctionTypes}}), consists of the following \fp s in Fig.\ref{fig:East-adl model}: {\gt{System}} function type contains four \fp s, i.e.,  the {\gt{Camera}} captures sign images and relays the images to {\gt{SignRecognition}} periodically. {\gt{Sign Recognition}} analyzes each frame of the detected images and computes the desired images (sign types). {\gt{Controller}} determines how the speed of the vehicle is adjusted based on the sign types and the current speed of the vehicle. {\gt{VehicleDynamic}} specifies the kinematics behaviors of the vehicle. {\gt{Environment}} function type consists of three \fp s, i.e., the information of traffic signs, random obstacles, and speed changes caused by environmental influences described in {\gt{TrafficSign}}, {\gt{Obstacle}}, and  {\gt{Speed}} \fp s respectively.

We consider the {\gt{Periodic, Execution, End-to-End, Synchronization, Sporadic}}, and {\gt{Comparison}} timing constraints on top of the \av\ \ed\ model, which are sufficient to capture the constraints described in Fig.\ref{fig:East-adl model}. Furthermore, we extend \ed/\tdl\ with an {\gt{Exclusion}} timing constraint that integrates relevant concepts from the \ccsl\ constraint, i.e., two events cannot occur simultaneously (R27 -- R31).

\vspace{0.1in}
\noindent R1. The camera must capture an image every 50ms. In other words, a {\gt{Periodic}}
acquisition of {\gt{Camera}} must be carried out every 50ms.

\noindent R2. The captured image must be recognized by an AV every 200ms, i.e., a {\gt{Periodic}} constraint on {\gt{SignRecognition}} \fp.

\noindent  R3. The obstacle must be detected by an AV every 40ms, i.e., the {\gt{Periodic}} timing constraint is applied on the input port of {\gt{Controller}}.

\noindent R4. The speed of AV is updated periodically with the period of 30ms.

\noindent R5. The detected image must be computed within [100, 150]ms in order to generate the desired sign type, {\gt{SignRecognition}} must complete its execution within [100, 150]ms.

\noindent R6. {\gt{Camera}} sends out captured images within [20, 30]ms to {\gt{Controller}}, i.e., the execution time of {\gt{Camera}} should be between 20 and 30ms.

\noindent R7. If an obstacle is detected, {\gt{Controller}} must send out a ``brake request'' signal to {\gt{VehicleDynamic}} in order to stop AV within [100, 150]ms, i.e., the execution time of {\gt{Controller}} should be in the range of [100, 150].

\noindent R8. After {\gt{VehicleDynamic}} receives a command/request from {\gt{Controller}}, the speed of AV should be updated within [50, 100]ms, i.e., the {\gt{Execution}} timing constraint applied on {\gt{VehicleDynamic}} is within [50, 100]ms.

\noindent R9. If the mode of AV switches to ``emergency stop'' due to a certain obstacle, it should not revert back to ``automatic running'' mode within a specific time period. It is interpreted as a {\gt{Sporadic}} constraint, i.e., the mode of AV is changed to {\gt{Stop}} because of the encounter with an obstacle, it should not revert back to {\gt{Run}} mode within 500ms.

\noindent R10. If the mode of AV switches to ``emergency stop'' due to a certain obstacle, it should not revert back to ``accelerate '' mode within a specific time period. It is interpreted as a {\gt{Sporadic}} constraint, i.e., the mode of AV is changed to {\gt{Stop}} because of  the encounter with an obstacle, it should not revert back to {\gt{accelerate }} mode within 500ms.

\noindent R11. If the mode of AV switches to ``emergency stop'' due to a certain obstacle, it should not revert back to ``turn left'' mode within a specific time period. It is interpreted as a {\gt{Sporadic}} constraint, i.e., the mode of AV is changed to {\gt{Stop}} because of the encounter with an obstacle, it should not revert back to {\gt{turn left}} mode within 500ms.

\noindent  R12. If the mode of AV switches to ``emergency stop'' due to the certain obstacle, it should not revert back to ``turn right'' mode within a specific time period. It is interpreted as a {\gt{Sporadic}} constraint, i.e., the mode of AV is changed to {\gt{Stop}} because of the encounter with an obstacle, it should not revert back to {\gt{turn right}} mode within 500ms.

\noindent  R13. The necessary information from environment must be arrived to {\gt{Controller}} within 40ms, e.g., all the input signals arriving on the {\gt{speed}}, {\gt{signType}}, {\gt{direct}}, {\gt{gear}} and {\gt{torque}} ports of {\gt{Controller}} must be within a given time window, i.e., the tolerated maximum constraint is 40ms. It is called {\gt{Input Synchronization}} constraint.

\noindent  R14. Once the execution of {\gt{Controller}} is completed, it sends out the computed signals/values to {\gt{VehicleDynamic}} within 30ms, e.g., all the ouput signals leaving via {\gt{reqTorq}}, {\gt{reqDirect}}, {\gt{reqGear}}, {\gt{reqBrake}} ports of {\gt{Controller}} must be within a given time window, i.e., the tolerated maximum constraint is 30ms. It is called {\gt{Output Synchronization}} constraint.

\noindent  R15. The necessary information from {\gt{Controller}} must be arrived to {\gt{VehicleDyna}}-{\gt{mic}} within 30ms. The {\gt{Input Synchronization}} constraint applied on the input ports of {\gt{VehicleDynamic}} ({\gt{reqTorq}}, {\gt{reqDirect}}, {\gt{reqGear}}, {\gt{reqBrake}}) should be 30ms.

\noindent  R16. Once {\gt{VehicleDynamic}} completes its execution, the information of AV must be updated within 40ms. The {\gt{Output Synchronization}} constraint applied on the output ports of {\gt{VehicleDynamic}} ({\gt{speed}}, {\gt{direct}}, {\gt{gear}}, {\gt{torque}}) should be 40ms.

\noindent  R17. When a traffic sign is recognized, the speed of AV should be updated within [150, 250]ms. An {\gt{End-to-End}} constraint on {\gt{Controller}} and {\gt{VehicleDy}}-\gt{{namic}}, i.e., the time interval from the input of {\gt{Controller}} to the output of {\gt{VehicleDynamic}} must be within a certain time.

\noindent  R18. When {\gt{Camera}} is triggered, the computation of image processing based on the traffic signs captured by {\gt{Camera}} must be finished within [120,  180]ms, i.e., the {\gt{End-to-End}} timing constraint applied on {\gt{Camera}} and  {\gt{SignRecognition}} should be between 120ms and 180ms.

\noindent  R19. The time interval between {\gt{Camera}} capturing an image of traffic sign and the status of AV (i.e., speed, direction etc.) being updated according to the recognized sign type should be within [270 , 430]ms,  {\gt{End-to-End}} timing constraint measured from the input of  {\gt{Camera}} to the output of {\gt{VehicleDynamic}} should be between 270 and 430ms.

\noindent  R20. When a ``left turn'' sign is recognized, AV must turn left within 500ms, i.e.,  a {\gt{End-to-End}} timing constraint applied on the events {\gt{DetectLeftSign}} and {\gt{StartTurn}}- \gt{Left} is 500ms.

\noindent  R21. When a ``right turn'' sign is recognized, AV must turn right within 500ms, i.e.,  a {\gt{End-to-End}} timing constraint applied on the events {\gt{DetectRightSign}} and {\gt{StartTurnRight}} is 500ms.

\noindent  R22. When a ``stop'' sign is recognized, AV must start to brake within 200ms, i.e., a {\gt{End-to-End}} timing constraint applied on the events {\gt{DetectStopSign}} and {\gt{StartBrake}} is 20ms.

\noindent  R23. When a ``stop'' sign is recognized, AV must be stop completely within 3000ms, i.e., a {\gt{End-to-End}} timing constraint applied on the events {\gt{DetectStopSign}} and {\gt{Stop}} is 3000ms.

\noindent  R24 The execution time interval between {\gt{Controller}} and {\gt{VehicleDynamic}} is less than or equal to the sum of the worst case execution time interval of each \fp.

\noindent  R25. The execution time interval between {\gt{Camera}} and {\gt{SignRecognition}} is less than or equal to the sum of the worst case execution time interval of each \fp.

\noindent  R26. The execution time interval between {\gt{Camera}} and {\gt{VehicleDynamic}}  is less than or equal to the sum of the worst case execution time interval of each \fp.

\noindent  R27. While AV turns left, the ``turning right'' mode should not be activated. The events of turning left and right considered as exclusive and specified as an {\gt{Exclusion}} constraint.

\noindent  R28. While AV is braking, the ``accelerate'' mode should not be activated. The events of braking and accelerating are considered as exclusive and specified as an {\gt{Exclusion}} constraint.

\noindent  R29. When AV is in the ``emergency'' mode because of encountering an obstacle, ``turning  left'' mode must not be activated, i.e., the events of handling ``emergency'' and ``turning left'' are exclusive. It is specified as an {\gt{Exclusion}} constraint.

\noindent  R30. When AV is in the ``emergency'' mode because of encountering an obstacle, ``turning right'' mode must not be activated, i.e., the events of handling ``emergency'' and ``turning right'' are exclusive. It is specified as an {\gt{Exclusion}} constraint.

\noindent  R31. When AV is in the ``emergency'' mode because of encountering an obstacle, ``accelerating'' mode must not be activated, i.e., the events of handling ``emergency'' and ``accelerating'' are exclusive. It is specified as an {\gt{Exclusion}} constraint.

{\gt{Delay}} constraint gives duration bounds (minimum and maximum)
between two events \emph{source} and \emph{target}.  This is specified using \emph{lower, upper}
values given as either {\gt{Execution}} constraints (R5 -- R8) or {\gt{End-to-End}} constraints (R17 -- R23). {\gt{Synchronization}} constraint describes how tightly the
occurrences of a group of events follow each other. All events must
occur within a sliding window, specified by the \emph{tolerance}
attribute, i.e., the maximum time interval allowed between events (R13 -- R16). {\gt{Periodic}} constraint  states that the period of successive
occurrences of a single event must have a time interval (R1 -- R4). {\gt{Sporadic}} constraint states that \emph{events} can arrive at arbitrary points in time, but with
defined minimum inter-arrival times between two consecutive occurrences (R9 -- R12). {\gt{Comparison}} constraint delimits that two consecutive occurrences of an event should have a minimum inter-arrival time (R24 -- R26). {\gt{Exclusion}} constraint states that two events must not occur at the same time (R27 -- R31). Those timing constraints are formally specified (seen as {\gt{Spec. R. IDs}} in Fig.2) using clock \emph{relation} and \emph{expression} in the context of WH then verified utilizing probabilistic analysis techniques that are described further in the following sections.

\chapter{Probabilistic Extension of \emph{Relations} in CCSL}
\label{sec:prccsl}
To perform the formal specification and probabilistic verification of \ed\ timing constraints (R1 -- R31 in Sec 3.), \ccsl\ \emph{relations} are augmented with probabilistic properties, called Pr\ccsl,  based on WH \cite{Bernat2001Weakly}. To describe the bound on the number of allowed constraint violations in WH, we extend \ccsl\ \emph{relations} with a probabilistic parameter $p$, where  $p$ is the probability threshold. Pr\ccsl\ is satisfied if and only if the probability of \emph{relation} constraint being satisfied is greater than or equal to $p$.
\begin{mydef} [\textbf{PrCCSL}]
Let $c1$, $c2$ and $\mathcal{M}$ be two logical clocks and a system model.
The probabilistic extension of relation constraints, denoted  $c1\textcolor{red}{\sim_{p}}c2$, is satisfied if the following condition holds:
\[\mathcal{M} \vDash c1\textcolor{red}{\sim_{p}}c2 \Longleftrightarrow {Pr}(c1 \textcolor{red}{\sim} c2) \geq p\]
where \textcolor{red}{$\sim$} $\in \{\textcolor{red}{\subseteq, \equiv, \prec, \preceq, \#} \}$,  ${Pr}(c1 \textcolor{red}{\sim} c2)$ is the probability of the relation $c1 \textcolor{red}{\sim} c2$ being satisfied, and $p \in [0,1]$ is the probability threshold.
\end{mydef}

\noindent ${Pr}(c1 \textcolor{red}{\sim} c2)$ is calculated based on clock ticks: ${Pr}(c1 \textcolor{red}{\sim} c2) = \frac{m}{k}$, where $k$ is the total number of ticks and $m$ is a number of  ticks satisfying the clock relation $c1 \textcolor{red}{\sim} c2$.

\begin{mydef} [\textbf{Tick and History}]
For $c \in C$, the tick of $c$ is indicated by a function t$_c$:  $\mathbb{N} \rightarrow \{0,1\}$. For $i \in \mathbb{N}$, t$_c$($i$) is a boolean variable that indicates whether $c$ ticks at the $i^{th}$ step, which is defined as: if \  $c$\ ticks\ at\ step\ $i$, t$_c$($i$) = 1; otherwise \ t$_c$($i$)\ =\ 0. The history of \emph{c} is a function h$_c$: $\mathbb{N} \rightarrow \mathbb{N}$. h$_c$($i$) that represents the number of ticks of \emph{c} that have been fired prior to the $i^{th}$ step, which can be defined as: (1) $h_c(0) = 0$; (2) $\forall\ i \in \mathbb{N^+},\  t_c(i)=0 \ \Longrightarrow \ h_c(i+1) = h_c(i)$; (3) $\forall\ i \in \mathbb{N^+}, \ t_c(i)=1 \ \Longrightarrow \ h_c(i+1) = h_c(i)\ +\ 1$.
\end{mydef}

The five \ccsl\ \emph{relations}, {\gt{Subclock}}, {\gt{Coincidence}}, {\gt{Exclusion}},  {\gt{Causality}} and {\gt{Precedence}}, are considered and the related probabilistic extensions are defined.

\begin{mydef}[\textbf{{Probabilistic Subclock}}] The probability of subclock relation
 between $c1$ and $c2$, denoted $c1\textcolor{red}{\subseteq_{p}}c2$, is satisfied if the following conditions hold:

\[\mathcal{M} \vDash c1\textcolor{red}{\subseteq_{p}}c2
\Longleftrightarrow {Pr}(c1 \textcolor{red}{\subseteq} c2) \geq p\]
where ${Pr}(c1 \textcolor{red}{\subseteq} c2)  =  \frac{m}{k}$,
${k} = \sum\limits_{i=0}^{n} t_{c1}(i)$,
${m} = \sum\limits_{i=0}^{n} \{t_{c1}(i) \wedge (t_{c1}(i) \Longrightarrow t_{c2}(i))\}$
\end{mydef}

\begin{figure}
  \centering
  \includegraphics[width=4.4in]{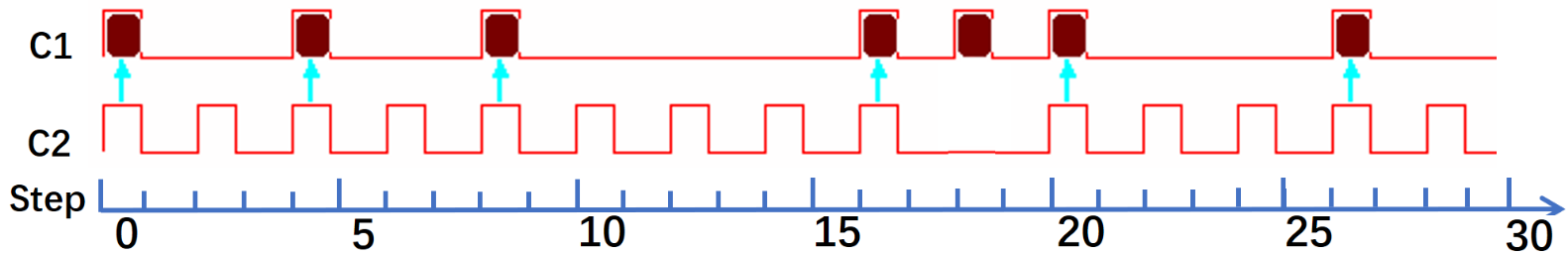}
  \caption{Example of subclock relation}
  \label{fig:subclk}
\end{figure}

\noindent $n$ refers to the simulation bound (number of steps of an execution). $k$ is the total number of ticks of the subclock $c1$ during the execution. $m$ is the number of ticks of $c1$ satisfying the {\gt{subclock}} relation. A tick of the subclock $c1$ satisfies the relation if at the step it occurs, its superclock $c2$ ticks. An example is shown in Fig. \ref{fig:subclk}: among the 30 steps, $c1$ ticks seven times, and six of them (denoted by the arrows) satisfy {\gt{subclock}} relation. In this case, $n$=30, $k$ = 7 and $m$ = 6.

{\gt{Coincidence}} relation states that two clocks should tick at the same step. i.e., they are subclocks of each other.
\begin{mydef}[\textbf{Probabilistic Coincidence}]  The probability of coincidence relation between $c1$ and $c2$, denoted
$c1\textcolor{red}{\equiv_{ p}}c2$, is satisfied if the following conditions hold:
\[\mathcal{M} \vDash c1\textcolor{red}{\equiv_{ p}}c2 \Longleftrightarrow {Pr}(c1 \textcolor{red}{\equiv} c2) \geq p\]
where ${Pr}(c1 \textcolor{red}{\equiv} c2)  = \frac{{m}}{{k}}$,
${k} =\sum\limits_{i=0}^{n} \{t_{c1}(i) \vee t_{c2}(i)\}$,
${m} = \sum\limits_{i=0}^{n} \{t_{c1}(i) \wedge t_{c2}(i)\}$
\end{mydef}

\noindent ${Pr}(c1{ \textcolor{red}{\equiv}}c2)$ represents the probability of the instants $c1$ that are coincident with the instants of $c2$.
{\gt{Coincidence}} relation is bidirectional, which means that $c1$ and $c2$ are equivalent in the relation. In this case, $k$ is the total number of steps at which either $c1$ or $c2$ ticks. $m$ is the number of ticks of steps at which {\gt{coincidence}} relation is satisfied, i.e., the steps at which both $c1$ and $c2$ tick.

The inverse of {\gt{coincidence}} relation, called {\gt{exclusion}}, hinders two clocks from ticking simultaneously.

\begin{mydef}[\textbf{Probabilistic Exclusion}] The probability of exclusion relation between $c1$ and $c2$, denoted
$c1\textcolor{red}{\#_{ p}}c2$, is satisfied if the following conditions hold:
\[\mathcal{M} \vDash c1\textcolor{red}{\#_{ p}}c2 \Longleftrightarrow {Pr}(c1 \textcolor{red}{\#} c2) \geq  p,\ \
where\ \ {Pr}(c1 \textcolor{red}{\#} c2)  = \frac{{m}}{{k}}, \]
\[{k} = \sum\limits_{i=0}^{n} \{t_{c1}(i) \vee t_{c2}(i)\},\]
\[{m} = \sum\limits_{i=0}^{n} \{(t_{c1}(i) \wedge \neg t_{c2}(i)) \vee (\neg t_{c1}(i) \wedge t_{c2}(i))\}\]
\end{mydef}

\noindent $k$ is the total number of steps at which either $c1$ or $c2$ ticks. $m$ indicates the number of steps at which {\gt{exclusion}} relation is satisfied, i.e., the steps at which only one of the two clocks ticks.

The probabilistic extension of {\gt{causality}} and {\gt{precedence}} relations are defined based on the history of the clocks. Recall that $h_{c1}(i)$ ($h_{c2}(i)$) indicates how many times $c1$ ($c2$) has ticked before the step $i$.
If the history of $c1$ is greater than the one of $c2$ at the same step, we say that $c1$ runs faster than $c2$ at that step.
{\gt{Causality}} relation specifies that an event causes another one, i.e., the effect cannot occur if the cause has not.

\begin{mydef}[\textbf{Probabilistic Causality}] The probabilistic causality
relation between $c1$ and $c2$ ($c1$ is the cause and $c2$ is the effect), denoted,
$c1\textcolor{red}{\preceq_{ p}}c2$, is satisfied if the following conditions hold:

\[\mathcal{M} \vDash c1\textcolor{red}{\preceq_{ p}}c2 \Longleftrightarrow {Pr}(c1 \textcolor{red}{\preceq} c2) \geq p\]
where ${Pr}(c1 \textcolor{red}{\preceq} c2)  = \frac{{m}}{{k}}$,
${k} = \sum\limits_{i=0}^{n} t_{c1}(i)$,
${m} = \sum\limits_{i=0}^{n} \{t_{c1}(i) \wedge  h_{c1}(i) \geq h_{c2}(i)\}$
\end{mydef}

\noindent $k$ is the total number of ticks of $c1$.  $m$ is the number of ticks of $c1$ satisfying {\gt{causality}} relation.
A tick of $c1$ satisfies  {\gt{causality}} relation if $c2$ does not occur prior to $c1$, i.e., the history of $c2$ is less than or equal to the history of $c1$ at the current step.

The strict {\gt{causality}}, called {\gt{precedence}}, constrains that one clock must always run faster than the other.

\begin{mydef}[\textbf{Probabilistic Precedence}] The probabilistic precedence relation between $c1$ and $c2$, denoted,
$c1\textcolor{red}{\prec_{ p}}c2$, is satisfied if the following conditions hold:
\[\mathcal{M} \vDash c1\textcolor{red}{\prec_{ p}}c2 \Longleftrightarrow {Pr}(c1 \textcolor{red}{\prec} c2) \geq p,\ \ where\]
\[{Pr}(c1 \textcolor{red}{\prec} c2)  = \frac{{m}}{{k}},\ \
{k} = \sum\limits_{i=0}^{n} t_{c1}(i),\]
\[
{m} = \sum\limits_{i=0}^{n} \underbrace{t_{c1}(i) \wedge h_{c1}(i)\geq h_{c2}(i)}_\text{(1)} \wedge \underbrace{(h_{c1}(i) = h_{c2}(i) \Longrightarrow \neg t_{c2}(i)}_\text{(2)})
\]
\end{mydef}

\noindent $k$ indicates the total number of ticks of $c1$. $m$ is the number of ticks of $c1$ satisfying {\gt{precedence}} and  holding the two conditions: (1) the history of $c1$ is greater than or equal to the history of $c2$ at the same step; (2) $c1$ and $c2$ must not be  coincident, i.e., when the history of $c1$ and $c2$ are equal, $c2$ must not tick.

\chapter{Specification of Timing Constraints in PrCCSL}
\label{sec: specification}
To describe the property that a timing constraint is satisfied with the probability greater than or equal to a given threshold,  \ccsl\ and its extension Pr\ccsl\ are employed to capture the semantics of probabilistic timing constraints in the context of WH.
Below, we show how \ed\ timing constraints, including {\gt{Execution}}, {\gt{Periodic}}, {\gt{End-to-End}}, {\gt{Sporadic}}, {\gt{Synchronization}}, {\gt{Exclusion}} and {\gt{Comparison}} timing constraints, can be specified in \ccsl/Pr\ccsl. In the system, events are represented as clocks. The ticks of clocks correspond to the occurrences of the events.

\vspace{0.1in}
\noindent \textbf{Periodic} timing constraints (R1 -- R4) can be specified using {\gt{PeriodicOn}} \emph{expression} and {\gt{probabilistic coincident}} \emph{relation}. For example, R1 states that the camera must be triggered periodically with a period 50ms. We first construct a periodic clock $prd\_50$ which ticks every $50$ ticks of $ms$ (the universal clock). Then the property that the {\gt{periodic}} timing constraint is satisfied with probability no less than the threshold $p$ can be interpreted as the probabilistic coincidence \emph{relation} between $cmrTrig$ (the event that {\gt{Camera}} \fp\ being triggered) and $prd\_50$. The corresponding specification is given below, where $\triangleq$ means ``is defined as'':
\begin{equation}
    prd\_50\ \triangleq\ \gt{PeriodicOn}\ ms\ \gt{period}\ 50
	\end{equation}
	\begin{equation}
	{cmrTrig}\ \textcolor{red}{\equiv_p}\ prd\_50
	\end{equation}
By combining (1) and (2), we can obtain the the specification of R1:
	\begin{equation}
	{cmrTrig}\ \textcolor{red}{\equiv_p}\ \{\gt{PeriodicOn}\ ms\ \gt{period}\ 50\}
	\end{equation}
In similar, the \ccsl/Pr\ccsl\ specification of R3 -- R4 can be derived:
	\begin{equation}
	\textbf{R3}:\ {obsDetect}\ \textcolor{red}{\equiv_p}\ \{\gt{PeriodicOn}\ ms\ \gt{period}\ 40\}
	\end{equation}
	\begin{equation}
	\textbf{R4}:\ {spUpdate}\ \textcolor{red}{\equiv_p}\ \{\gt{PeriodicOn}\ ms\ \gt{period}\ 30\}
	\end{equation}
where $signTrig$ is the event/clock that {\gt{SignRecognition}} \fp\ will be triggered, $obsDetect$ represents the event that the object detection is activated by the vehicle and $spUpdate$ denotes the event that the speed is updated (i.e., recieved by  {\gt{Controller}}) from the environment.

\noindent Since the $period$ attribute of the periodic timing constraint R2 is 200ms, which is a integral multiple of the $period$ of R1,  R2 can be interpreted as a {\gt{subclock}} relation, i.e., the event $signTrig$ should be a \emph{subclock} of $cmrTrig$. The specification is given below:
	\begin{equation}
	\ {signTrig}\ \textcolor{red}{\subseteq_p}\ cmrTrig
	\end{equation}
\vspace{0.1in}
\noindent \textbf{{Execution}} timing constraints (R5 -- R8) can be specified using {\gt{DelayFor}} \emph{expression} and {\gt{probabilistic causality}} \emph{relation}. To specify R5, which states that the {\gt{SignRecognition}} \fp\ must finish execution within [100, 150]ms, i.e., the interval measured from the input event of the \fp\ (i.e., the event that the image is received by the \fp, denoted $imIn$) to the output event of the \fp\ (denoted $signOut$) must have a minimum value 100 and a maximum value 150. We divide this property into two sub-properties:
R5(1). The time duration between $imIn$ and  $signOut$ should be greater than 100ms.
R5(2). The time duration between $imIn$ and  $signOut$ should be less than 150ms.
To specify property R5(1), we first construct a new clock $imIn\_dly100$ by delaying $imIn$ (the input event of {\gt{SignRecognition}}) for 100ms. To check whether R5(1) is satisfied within a probability threshold is to verify whether the {\gt{probabilistic causality}} between $imIn\_dly100$ and $signOut$ is valid. The specification of R5(1) is given below:
	\begin{equation}
    imIn\_dly100\ \triangleq\ imIn\ \gt{DelayFor}\  100\ \gt{on}\ ms
	\end{equation}
	\begin{equation}
	{imIn\_dly100}\ \textcolor{red}{\preceq_p}\ signOut
	\end{equation}
By combining (7) and (8), we can obtain the the specification of R5(1):
	\begin{equation}
	\{ {imIn}\ \gt{DelayFor}\ 100\ \gt{on}\ ms\}\ \textcolor{red}{\preceq_p}\ signOut
	\end{equation}

Similarly, to specify property R5(2), a new clock $imIn\_dly150$ is generated by delaying $imIn$ for 150 ticks on $ms$. Afterwards, the property that R5(2) is satisfied with a probability greater than or equal to $p$ relies on whether the {\gt{probabilistic causality}} \emph{relation} between $imIn\_dly150$ and $signOut$ is satisfied. The specification is illustrated as follows:
	\begin{equation}
    imIn\_dly150\ \triangleq\ imIn\ \gt{DelayFor}\  150\ \gt{on}\ ms
	\end{equation}
	\begin{equation}
	{signOut}\ \textcolor{red}{\preceq_p}\ mIn\_dly150
	\end{equation}
By combining (10) and (11), we can obtain the the specification of R5(2):
	\begin{equation}
	{signOut}\ \textcolor{red}{\preceq_p}\ \{ {imIn}\ \gt{DelayFor}\ 150\ \gt{on}\ ms\}
	\end{equation}
    Analogously, the \ccsl/Pr\ccsl\ specification of R6 -- R8 can be derived:\\
	\begin{equation}
    \begin{split}
     \textbf{R6}:\ \{ {cmrTrig}\ \gt{DelayFor}\ 20\ \gt{on}\ ms\}\ \textcolor{red}{\preceq_p}\ cmrOut\\
     {cmrOut}\ \textcolor{red}{\preceq_p}\ \{ {cmrTrig}\ \gt{DelayFor}\ 30\ \gt{on}\ ms\}
     \end{split}
	\end{equation}
	\begin{equation}
    \begin{split}
    \textbf{R7}:\ \{ {ctrlIn}\ \gt{DelayFor}\ 100\ \gt{on}\ ms\}\ \textcolor{red}{\preceq_p}\ ctrlOut\\
     {ctrlOut}\ \textcolor{red}{\preceq_p}\ \{ {ctrlIn}\ \gt{DelayFor}\ 150\ \gt{on}\ ms\}
     \end{split}
	\end{equation}
	\begin{equation}
    \begin{split}
    \textbf{R8}:\ \{ {vdIn}\ \gt{DelayFor}\ 50\ \gt{on}\ ms\}\ \textcolor{red}{\preceq_p}\ vdOut\\
     {vdOut}\ \textcolor{red}{\preceq_p}\ \{ {vdIn}\ \gt{DelayFor}\ 100\ \gt{on}\ ms\}
     \end{split}
	\end{equation}
where $cmrTrig$ is the event that the {\gt{Camera}} \fp\ is triggered, $cmrOut$ represents the event that the captured image is sent out. $ctrlIn$ ($ctrlOut$) represents the input (resp. output)  of  {\gt{Controller}} \fp. $vdIn$ ($vdOut$) represents the input (resp. output)  of  {\gt{VehicleDynamic}} \fp.

\vspace{0.1in}
\noindent \textbf{Sporadic} timing constraints (R9 -- R12) can be specified using {\gt{DelayFor}} \emph{expression} and {\gt{probabilistic precedence}} \emph{relation}. For instance, R9 states that there should be a minimum delay between the event $veRun$ (the event that the vehicle is in the \emph{Run} mode) and the event $obstc$ (the event that the vehicle detects an obstacle), which is specified as 500ms.
To specify R9, we first build a new clock $obstc\_dly500$ by delaying $obstc$ for 500 ticks of $ms$. We then check whether the {\gt{probabilistic precedence}} relation between $obst\_dly500$ and $veRun$:
	\begin{equation}
    obstc\_dly500\ \triangleq\ obstc\ \gt{DelayFor}\  500\ \gt{on}\ ms
	\end{equation}
	\begin{equation}
	{ obstc\_dly500}\ \textcolor{red}{\prec_p}\ veRun
	\end{equation}
By combining (16) and (17), we can obtain the the specification of R9:
	\begin{equation}
	\{obstc\ \gt{DelayFor}\  500\ \gt{on}\ ms\}\ \textcolor{red}{\prec_p}\ veRun
	\end{equation}
    Analogously, the \ccsl/Pr\ccsl\ specification of R10 -- R12 can be derived:\\
	\begin{equation}
	\textbf{R10}:\ \{obstc\ \gt{DelayFor}\  500\ \gt{on}\ ms\}\ \textcolor{red}{\prec_p}\ veAcc
	\end{equation}
	\begin{equation}
	\textbf{R11}:\ \{obstc\ \gt{DelayFor}\  500\ \gt{on}\ ms\}\ \textcolor{red}{\prec_p}\ tLeft
	\end{equation}
	\begin{equation}
	\textbf{R12}:\ \{obstc\ \gt{DelayFor}\  500\ \gt{on}\ ms\}\ \textcolor{red}{\prec_p}\ tRight
	\end{equation}
where $veAcc$ is the event/clock that the vehicle is accelerating. $tLeft$ and $tRight$ represent the event that the vehicle transits from the \emph{emergency} stop mode to \emph{turn left} and \emph{turn right} mode respectively.

\vspace{0.1in}
\noindent \textbf{Synchronization} timing constraints (R13 -- R16) can be specified using {\gt{Infimum}} and {\gt{Supremum}} \emph{expression}, together with {\gt{probabilistic precedence}} \emph{relation}.
R13 states that the five input events must be detected by {\gt{Controller}}
within the maximum tolerated time, given as 40ms.
The synchronization timing constraint can be interpreted as: the time interval between the earliest/fastest and the latest/slowest event among the five input events, i.e.,  \emph{speed}, \emph{signType}, \emph{direct}, \emph{gear} and \emph{torque},  must not exceed 40ms.
To specify the constraints, {\gt{Infimum}} is utilized to express
the fastest event (denoted $inf_{ctrlIn}$) while {\gt{Supremum}} is utilized to specify the slowest event $sup_{ctrlIn}$. $sup_{ctrlIn}$ and $inf_{ctrlIn}$ are defined as:
\begin{equation}
sup_{ctrl}\ \triangleq\ {{\gt{Sup}}}({{\gt{Sup}}}(speed,\ signType),\ {{\gt{Sup}}}({\gt{Sup}}(direct,\ gear),\ toque))
\end{equation}
\begin{equation}
inf_{ctrl}\ \triangleq\ {\gt{Inf}}({\gt{Inf}}(speed,\ signType),\ {\gt{Inf}}({\gt{Inf}}(direct,\ gear),\ toque))
\end{equation}
where {\gt{Inf}}($c1$, $c2$) (resp. {\gt{Sup}}($c1$, $c2$)) is the {\gt{Infimum}} (resp. {\gt{Supremum}}) operator returns the
slowest (resp. fastest) clock faster (resp. slower) than $c1$ and $c2$.
Afterwards, we construct a new clock $inf_{ctrlIn}\_dly40$ that is the $inf_{ctrlIn}$ delayed for 40 ticks of $ms$, which is defined as:
	\begin{equation}
    inf_{ctrlIn}\_dly40\ \triangleq\ inf_{ctrlIn}\ \gt{DelayFor}\  40\ \gt{on}\ ms
	\end{equation}
Therefore, the {\gt{synchronization}} constraint R13 can be represented as the {\gt{probab}}-\gt{{ilistic causality}} \emph{relation} between $sup_{ctrlIn}$ and $inf_{ctrlIn}\_dly40$, given as the \ccsl/Pr\ccsl\ expression below:
	\begin{equation}
    sup_{ctrlIn}\ \textcolor{red}{\preceq_p}\  {inf_{ctrlIn}\_dly40}
	\end{equation}
By combining (24) and (25), we can obtain the the specification of R13:
	\begin{equation}
    sup_{ctrlIn}\ \textcolor{red}{\preceq_p}\  \{inf_{ctrlIn}\ \gt{DelayFor}\  40\ \gt{on}\ ms\}
	\end{equation}

\noindent In similar, the \ccsl/Pr\ccsl\ specification of R14 -- R16 can be derived.
For R14, we first construct the clocks that represent the fastest and slowest output event/clock among the four output events of {\gt{Controller}} \fp, i.e., $reqTorq$, $reqDirec$, $reqGear$ and $reqBrake$. Then the property that the synchronization constraint is satisfied with a probability greater than or equal to $p$ can be interpreted as a {\gt{probabilistic causality}} \emph{relation}:
	\begin{equation}
    \begin{split}
	\textbf{R14}:
    sup_{ctrlOut}\ \triangleq\ \gt{Sup}(\gt{Sup}(reqTorq,\ reqDirec),\ \gt{Sup}(reqGear,\ reqBrake))\\
    inf_{ctrlOut}\ \triangleq\ \gt{Inf}(\gt{Inf}(reqTorq,\ reqDirec),\ \gt{Inf}(reqGear,\ reqBrake))\\
    sup_{ctrlOut}\ \textcolor{red}{\preceq_p}\  \{inf_{ctrlOut}\ \gt{DelayFor}\  30\ \gt{on}\ ms\}
    \end{split}
    \end{equation}
For R15, we first construct the fastest and slowest input event/clock among the four input events of {\gt{VehicleDynamic}} , i.e., $reqTorq$, $reqDirec$, $reqGear$ and $reqBrake$. Then the property that the synchronization constraint is satisfied with a probability greater than or equal to $p$ can be interpreted as a {\gt{probabilistic causality}} \emph{relation}:
	\begin{equation}
    \begin{split}
	\textbf{R15}:
    sup_{vdIn}\ \triangleq\ \gt{Sup}(\gt{Sup}(reqTorq,\ reqDirec),\ \gt{Sup}(reqGear,\ reqBrake))\\
    inf_{vdIn}\ \triangleq\ \gt{Inf}(\gt{Inf}(reqTorq,\ reqDirec),\ \gt{Inf}(reqGear,\ reqBrake))\\
    sup_{vdIn}\ \textcolor{red}{\preceq_p}\  \{inf_{vdIn}\ \gt{DelayFor}\  40\ \gt{on}\ ms\}
    \end{split}
    \end{equation}
For R16, we first construct the fastest and slowest output event/clock among the four output events of {\gt{VehicleDynamic}} , i.e., $speed$, $direct$, $torque$ and $gear$. Then the property that the synchronization constraint is satisfied with a probability greater than or equal to $p$ can be interpreted as a {\gt{probabilistic causality}} \emph{relation}:
    \begin{equation}
    \begin{split}
	\textbf{R16}:
    sup_{vdOut}\ \triangleq\ \gt{Sup}(\gt{Sup}(speed,\ direct),\ \gt{Sup}(gear,\ torq))\\
    inf_{vdOut}\ \triangleq\ \gt{Inf}(\gt{Inf}(speed,\ direct),\ \gt{Inf}(gear,\ torque))\\
    sup_{vdOut}\ \textcolor{red}{\preceq_p}\  \{inf_{vdOut}\ \gt{DelayFor}\  40\ \gt{on}\ ms\}
    \end{split}
    \end{equation}

\vspace{0.1in}
\noindent \textbf{{End-to-End}} timing constraints (R17 -- R23) can be specified using {\gt{DelayFor}} \emph{expression} and {\gt{probabilistic precedence}} \emph{relation}.
To specify R17, which limits that the time duration measured from the instant of the occurrence of the event that {\gt{Controller}} \fp\ receive the traffic sign type information  (denoted as $signIn$),   to the occurrence of event that the speed is sent out from  {\gt{VehicleDynamic}} \fp\ (denoted as $spOut$) should be between 150 and 250ms.
We divide this property into two subproperties:
R17(1). The time duration between $signIn$ and  $spOut$ should be larger than 150ms.
R17(2). The time duration between $signIn$ and  $spOut$ should be less than 250ms.
To specify property R17(1), we first construct a new clock $signIn\_dly150$ by delaying $signIn$  for 150ms. To check whether R17(1) is satisfied within a probability threshold $p$ is to verify whether the {\gt{probabilistic precedence}} between $signIn\_dly150$ and $spOut$ is valid. The specification of R17(1) is given below:
	\begin{equation}
    signIn\_dly150\ \triangleq\ signIn\ \gt{DelayFor}\  150\ \gt{on}\ ms
	\end{equation}
	\begin{equation}
	{signIn\_dly150}\ \textcolor{red}{\prec_p}\ spOut
	\end{equation}
By combining (30) and (31), we can obtain the the specification of R17(1):
	\begin{equation}
	\{ {signIn}\ \gt{DelayFor}\ 150\ \gt{on}\ ms\}\ \textcolor{red}{\prec_p}\ spOut
	\end{equation}

\noindent Similarly, to specify property R17(2), a new clock $signIn\_dly250$ is generated by delaying $signIn$ for 250 ticks of $ms$. Afterwards, the property that R17(2) is satisfied with a probability greater than or equal to $p$ relies on whether the {\gt{probabilistic precedence}} \emph{relation} is satisfied. The specification is illustrated as follows:
	\begin{equation}
    signIn\_dly250\ \triangleq\ signIn\ \gt{DelayFor}\  250\ \gt{on}\ ms
	\end{equation}
	\begin{equation}
	{spOut}\ \textcolor{red}{\prec_p}\  signIn\_dly250
	\end{equation}
By combining (33) and (34), we can obtain the the specification of R17(2):
	\begin{equation}
	{spOut}\ \textcolor{red}{\prec_p}\ \{ {signIn}\ \gt{DelayFor}\ 250\ \gt{on}\ ms\}
	\end{equation}
In similar, the \ccsl/Pr\ccsl\ specification of R18 -- R23 can be derived:\\
	\begin{equation}
    \begin{split}
     \textbf{R18}:\ \{ {cmrTrig}\ \gt{DelayFor}\ 120\ \gt{on}\ ms\}\ \textcolor{red}{\prec_p}\ signOut\\
     {signOut}\ \textcolor{red}{\prec_p}\ \{ {cmrTrig}\ \gt{DelayFor}\ 180\ \gt{on}\ ms\}
     \end{split}
	\end{equation}
	\begin{equation}
    \begin{split}
    \textbf{R19}:\ \{ {cmrTrig}\ \gt{DelayFor}\ 270\ \gt{on}\ ms\}\ \textcolor{red}{\prec_p}\ spOut\\
     {spOut}\ \textcolor{red}{\prec_p}\ \{ {cmrTrig}\ \gt{DelayFor}\ 430\ \gt{on}\ ms\}
     \end{split}
	\end{equation}
	\begin{equation}
    \textbf{R20}:\ \{ {startTurnLeft}\ \textcolor{red}{\prec_p}\  DetectLeftSign\ \gt{DelayFor}\ 500\ \gt{on}\ ms\}
	\end{equation}
	\begin{equation}
    \textbf{R21}:\ \{ {startTurnRight}\ \textcolor{red}{\prec_p}\  DetectRightSign\ \gt{DelayFor}\ 500\ \gt{on}\ ms\}
	\end{equation}
	\begin{equation}
    \textbf{R22}:\ \{ {startBrake}\ \textcolor{red}{\prec_p}\  DetectStopSign\ \gt{DelayFor}\ 500\ \gt{on}\ ms\}
	\end{equation}
	\begin{equation}
    \textbf{R23}:\ \{{Stop}\ \textcolor{red}{\prec_p}\  DetectStopSign\ \gt{DelayFor}\ 3000\ \gt{on}\ ms\}
	\end{equation}

\vspace{0.1in}
\noindent \textbf{Comparison} timing constraints (R24 -- R26) can be specified using {\gt{DelayFor}} \emph{expression} and {\gt{probabilistic causality}} \emph{relation}. R24 states that the execution time interval from {\gt{Controller}} to {\gt{VehicleDynamic}} should be less than or equal to the sum of the worst case execution time of {\gt{Controller}} and {\gt{VehicleDy}}-\gt{{namic}}, denoted as \emph{W$_{ctrl}$} and \emph{W$_{vd}$} respectively.
To specify {\gt{comparison}} constraint, we first construct a new clock $signIn\_dly250$ by delaying $signIn$ for 250 ticks of $ms$. Afterwards, we generate another new clock $signIn\_dlysw$ that is the $signIn$ clock delayed for sum of the \emph{worst case execution time} of the two \fp s. The specification is illustrated as follows:
	\begin{equation}
    signIn\_dly250\ \triangleq\ signIn\ \gt{DelayFor}\  250\ \gt{on}\ ms
	\end{equation}
	\begin{equation}
    signIn\_dlysw\ \triangleq\ signIn\ \gt{DelayFor}\  (W_{ctrl}+W_{vd})\ \gt{on}\ ms
	\end{equation}

\noindent Therefore, the property that the probability of comparison constraint is satisfied should be greater than or equal to the threshold $p$ can be interpreted as a
 {\gt{probabilistic causality}} \emph{relation} between $signIn\_dly250$ and $ signIn\_dlysw$:
 	\begin{equation}
	 signIn\_dly250\ \textcolor{red}{\preceq_p}\ signIn\_dlysw
	\end{equation}
By combining (42), (43) and (44), we can obtain the the specification of R24:
	\begin{equation}
	\{signIn\ \gt{DelayFor}\  250\ \gt{on}\ ms\}\ \textcolor{red}{\preceq_p}\ \{signIn\ \gt{DelayFor}\  (W_{ctrl}+W_{vd})\ \gt{on}\ ms\}
	\end{equation}
\noindent Analogously, the \ccsl/Pr\ccsl\ specification of R25 and R26 can be derived:\\
    	\begin{equation}
    \begin{split}
	\textbf{R25}:\ \{cmrTrig\ \gt{DelayFor}\  180\ \gt{on}\ ms\}\ \textcolor{red}{\preceq_p}\ \\ \{cmrTrig\ \gt{DelayFor}\  (W_{cmr}+W_{sr})\ \gt{on}\ ms\}
    \end{split}
	\end{equation}
	\begin{equation}
    \begin{split}
	\textbf{R26}:\ \{cmrTrig\ \gt{DelayFor}\  430\ \gt{on}\ ms\}\ \textcolor{red}{\preceq_p}\ \\ \{cmrTrig\ \gt{DelayFor}\ (W_{cmr}+ W_{sr}+ W_{ctrl}+W_{vd})\ \gt{on}\ ms\}
    \end{split}
	\end{equation}
where $W_{cmr}$ and $W_{vd}$ represent the worst case execution time of {\gt{Camera}} and {\gt{SignRecognition}} respectively.

\vspace{0.1in}
\noindent \textbf{Exclusion} timing constraints (R27 -- R31) can be specified using {\gt{exclusion}} \emph{relation} directly. R27 states that the two events $turnLeft$ (the event that the vehicle is turning left) and $rightOn$ (the event that the turn right mode is activated) should be exclusive, which can be expressed as:
	\begin{equation}
    {turnLeft}\ \textcolor{red}{\#_p}\ rightOn
	\end{equation}
Analogously, the {\gt{exclusion}} timing constraints R28 -- R31 can be specified using {\gt{exclusion}} \emph{relation}:
	\begin{equation}
    \textbf{R27}:\ {turnLeft}\ \textcolor{red}{\#_p}\ rightOn
	\end{equation}
	\begin{equation}
    \textbf{R28}:\ {veAcc}\ \textcolor{red}{\#_p}\ veBrake
	\end{equation}
	\begin{equation}
    \textbf{R29}:\ {emgcy}\ \textcolor{red}{\#_p}\ turnLeft
	\end{equation}
	\begin{equation}
    \textbf{R30}:\ {emgcy}\ \textcolor{red}{\#_p}\ rightOn
	\end{equation}
	\begin{equation}
    \textbf{R31}:\ {emgcy}\ \textcolor{red}{\#_p}\ veAcc
	\end{equation}
where $emgcy$ is the event that the vehicle is in the \emph{emergency} mode, $veBrake$ and $veAcc$ represent the event that the vehicle is braking or accelerating, respectively.

\chapter{Translation of CCSL \& PrCCSL into SDV}
\label{sec:model}
In order to formally prove the \ed\ timing constraints (given in Sec. 3) using \sldv\ (SDV), we investigate how those constraints, specified in \ccsl\ \emph{expressions} and Pr\ccsl\ \emph{relations} ({\gt{Spec. R.ID}} in Fig. \ref{fig:East-adl model}), can be translated into \emph{Proof Objective Models} (POM).
\ccsl\ \emph{expressions} constructs new clocks and the \emph{relations} between the new clocks are specified using Pr\ccsl. We first provide strategies that represent \ccsl\ \emph{expressions} in \simu/\staf\ (S/S). We then present how the \ed\ timing constraints defined in Pr\ccsl\ can be translated into the corresponding POMs, which are integrated with the S/S models of \ccsl\ \emph{expressions}, based on the strategies.
\section{Mapping CCSL Expressions into S/S}
\begin{figure}[htbp]
  \centering
  \includegraphics[width=3in]{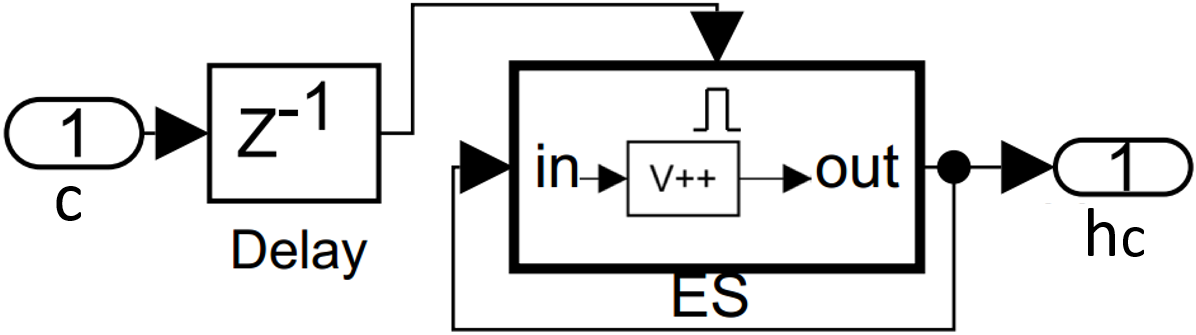}
  \caption{{$h_c$ = \gt{His(c)}}}
  \label{fig:his}
\end{figure}

We first describe how tick and history of \ccsl\ can be mapped to corresponding S/S models. Using the mapping, we show \ccsl\ \emph{expressions} can be modeled in S/S. A ``step'' (defined in Sec. \ref{sec:preli}) is represented as a sample time in \simu\ and set to 0.001 second.
The clock ticks are expressed as boolean variables (1 ``ticking'' or 0 ``non-ticking'') during simulation. The history of clock (expressed as integer) is increased as the clock ticks and interpreted as a function {\gt{His(c)}} in Fig.\ref{fig:his}:
Since h$_c$, the history of clock $c$, is determined by the value of $c$ at the immediate precedent step, a {\scriptsize $\ll$}Delay{\scriptsize $\gg$} block is employed to delay $c$ by one step. Whenever $c$ ticks at the prior step, {\scriptsize $\ll$}ES{\scriptsize $\gg$} is executed and increases h$_c$ by 1.

Based on the mapping patterns of tick and history, we present how {\gt{PeriodicOn}},  {\gt{DelayFor}},  {\gt{Infimum}} and  {\gt{Supremum}} \emph{expressions} can be represented as S/S models.

\begin{figure}[htbp]
  \centering
  \includegraphics[width=5in]{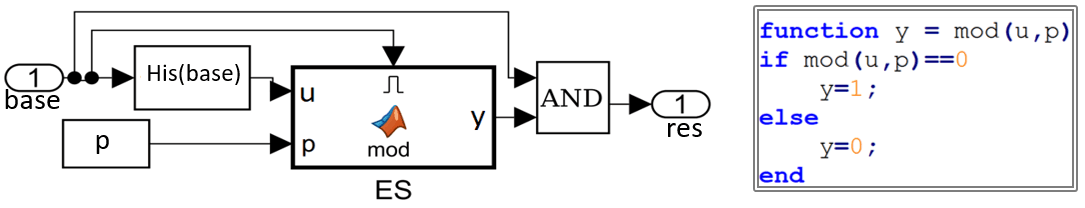}
  \caption{{\gt{$res\ \triangleq\ {\gt{PeriodicOn}}\ base\ {\gt{period}}\ p$}}}
  \label{fig:periodic}
\end{figure}

\vspace{0.1in}
\noindent\textbf{PeriodicOn}: $res \ \triangleq\ \ {\gt{PeriodicOn}}\ base\ {\gt{period}}\ p$, where $\triangleq$ means ``is defined as'',  builds a new clock $res$ based on $base$ clock and a \emph{period} parameter $p$, i.e., $res$ ticks at every $p^{th}$ tick of $base$. The \simu\ model of {\gt{PeriodicOn}} is illustrated in Fig.\ref{fig:periodic}: When \emph{base} ticks, the {\scriptsize $\ll$}Matlab Function{\scriptsize $\gg$} (code is shown in the box), embedded in the {\scriptsize $\ll$}ES{\scriptsize $\gg$} subsystem, is  triggered and checks if the history of the $base$, {\gt{His(base)}}, is an integral multiple of $p$. When \emph{base} ticks and its history equals to the integral multiple of $p$, $res$ ticks. The {{\gt{PeriodicOn}}} S/S model is employed for the translation of \ed\ {\gt{Periodic}} timing constraint (R1 -- R4 in Fig.\ref{fig:East-adl model}) into its POM in \sdv.

\begin{figure}[htbp]
  \centering
  \includegraphics[width=5.8in]{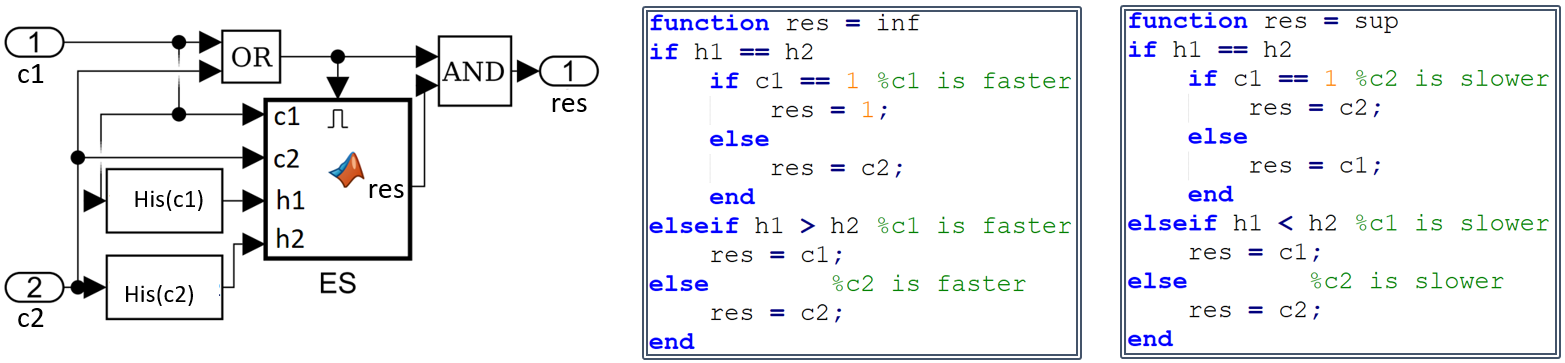}\\
  \caption{ $res\ \triangleq$ {\gt{Inf}}($c1$, $c2$) (rep. {\gt{Sup}}($c1$, $c2$))}
  \label{fig:infsup}
\end{figure}

\begin{figure}[htbp]
  \centering
  \includegraphics[width=5in]{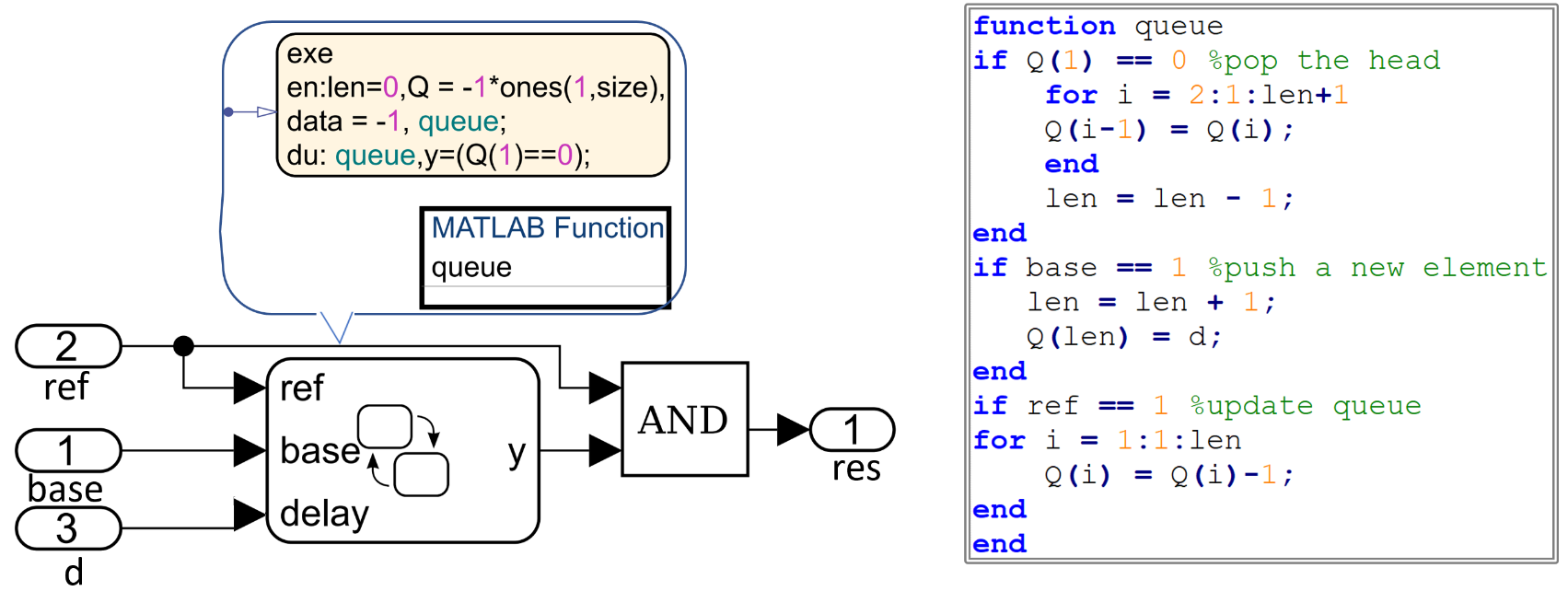}\\
  \caption{{\gt{$res \ \triangleq\ base\ {\gt{DelayFor}}\ d\ {\gt{on}}\ ref$}}}
  \label{fig:delay}
\end{figure}

\vspace{0.1in}
\noindent\textbf{Infimum (resp. Supremum)}: $res\ \triangleq$ {\gt{Inf}}($c1$, $c2$) (resp. {\gt{Sup}}($c1$, $c2$)), creates a new clock $res$, which is the slowest (resp. fastest) clock faster (resp. slower) than the two clocks, $c1$ and $c2$. In other words, $res$ ticks at the step whereby the faster (slower) clock between $c1$ and $c2$ ticks.
The \simu\ model of \gt{Infimum} (resp. \gt{Supremum}) is depicted in Fig.\ref{fig:infsup}.  When $c1$ or $c2$ ticks, the {\gt{inf}} (resp. {\gt{sup}}) function embedded in {\scriptsize $\ll$}ES{\scriptsize $\gg$} is executed and decides which clock is faster (resp. slower) than the other by comparing the history of $c1$ and $c2$ (h1 and h2). If the clock (either $c1$ or $c2$) ticking at the current step is the faster (resp. slower) clock, $res$ ticks. The {\gt{Infimum}} and {\gt{Supremum}} S/S models are utilized for the translation of \ed\ {\gt{Synchronization}} timing constraint (R13 -- R16 in Sec. \ref{sec:case-study}) into POM.
\vspace{0.1in}
\noindent\textbf{DelayFor}: $res \ \triangleq\ base\ {\gt{DelayFor}}\ d\ {\gt{on}}\ ref$,  constructs a new clock $res$ based on \emph{base} clock and \emph{reference} clock ($ref$), i.e., each time $base$ ticks, $res$  ticks at the $d^{th}$ tick of $ref$.
The \simu\ model of {\gt{DelayFor}} is shown in Fig. \ref{fig:delay}: A \staf\ chart is utilized to observe the ticks of $base$ and $ref$. A queue, \emph{Q}, whose enqueue/dequeue operation is implemented in the function \emph{queue}.  $y$ indicates whether $ref$ has ticked $d$ times since $base$ ticked.
When $base$ ticks ($base==1$), an element with value $d$ is enqueued, and each time $ref$ ticks, the value of the element is decreased by 1. After $d$ ticks of $ref$, the element becomes 0 and $y$ becomes true. An {\scriptsize $\ll$}And{\scriptsize $\gg$} block is applied to delimit that the tick of $res$ must coincide with the tick of $ref$ (i.e., $res$ is a \gt{subclock} of $ref$).
The {\gt{DelayFor}} S/S model is adapted to construct the POM models of \ed\ timing requirements
R5 -- R26 in Sec. \ref{sec:case-study}.

\section{Representation of PrCCSL in SDV}
We present how the translation of \ed\ timing constraints (specified in Pr\ccsl\ \emph{relations} and \ccsl\ \emph{expressions}) can be interpreted as POMs in the view point of analysis engine \sdv. Recall the definitions of Pr\ccsl\ in Sec. 4. A PrCCSL \emph{relation} is valid if the probability of a \emph{relation} $\phi$ being satisfied is greater than or equal to the given probability threshold $p$. It can be interpreted as a \emph{Hypothesis Testing} \cite{Reijsbergen2015On}: Decide whether $\mathcal{M}$ $\vDash$ Pr($\phi$)$\geq$ p (hypothesis H$_0$) against $\mathcal{M}$ $\vDash$ Pr($\phi$)$<$ p (alternative hypothesis H$_1$).

\begin{figure}[htbp]
\centering
  \subfigure[$signRecTrig\ \textcolor{red}{\subseteq_p}\ cTrig$]{
  \includegraphics[width=3.3in]{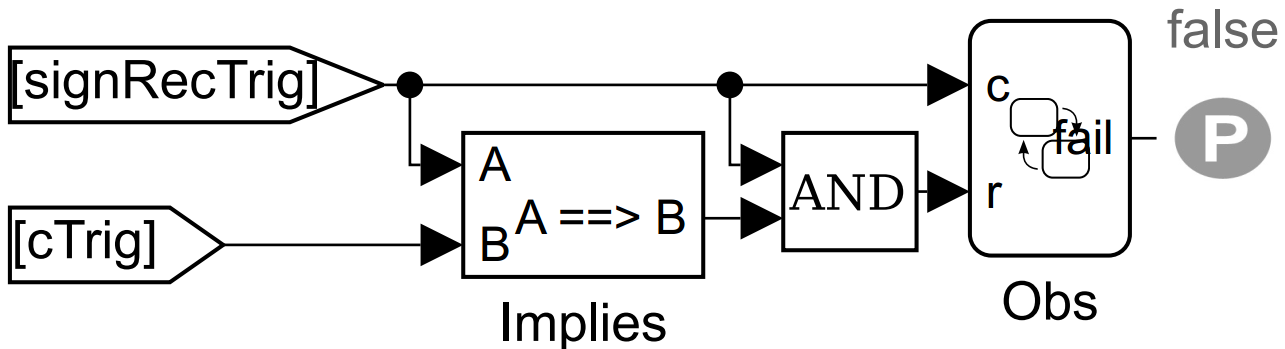}}
  \subfigure[$Obs$ Chart]{
  \includegraphics[width=3in]{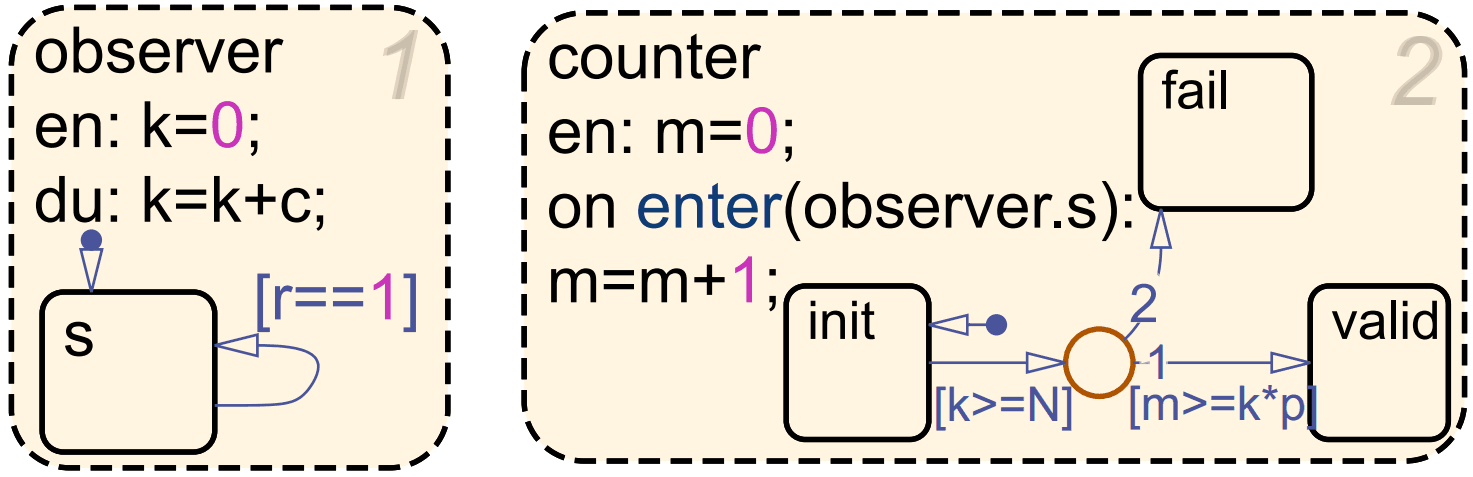}}
   \caption{POM of Probabilistic {\gt{Subclock}}}
\label{fig:pomsub}
\end{figure}

\begin{figure}[htbp]
\centering
  \subfigure[ $cTrig\ \textcolor{red}{\equiv_{p}}\ \{{\gt{PeriodicOn}}\ ms\ {\gt{period}}\ 50\}$]{
  \includegraphics[width=3.4in]{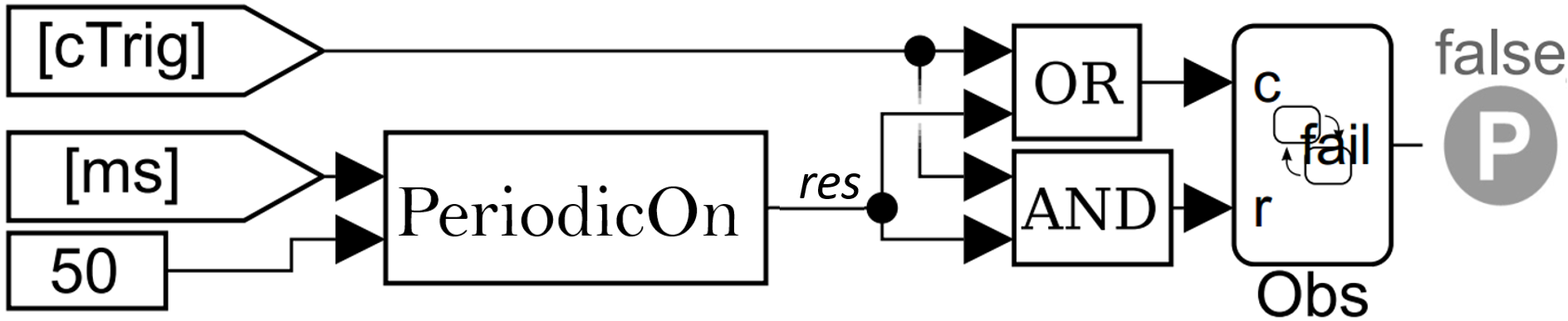}}
  \subfigure[$turnLeft\ \textcolor{red}{\#_p}\ rightOn$]{
  \includegraphics[width=3.2in]{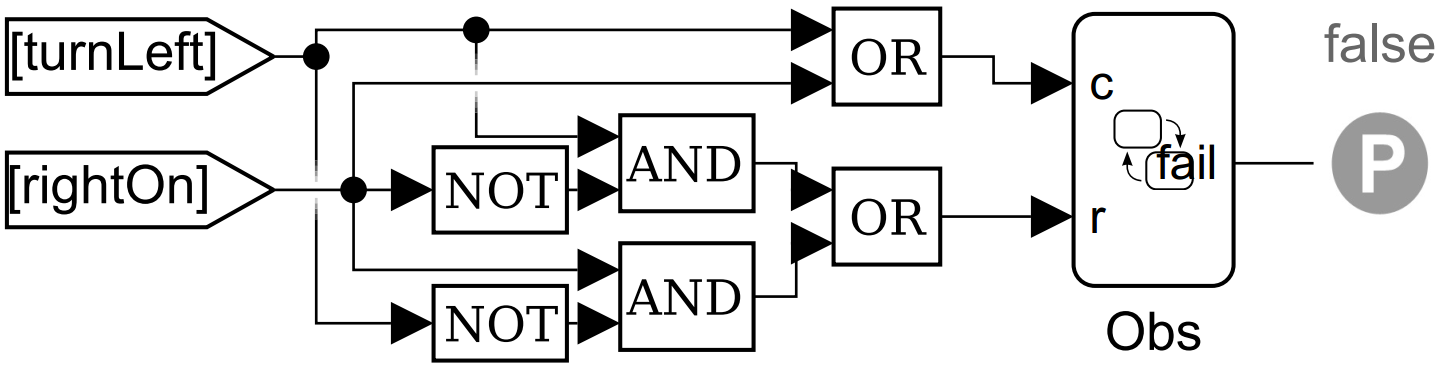}}
 % \subfigure[From and Goto]{
  %\includegraphics[width=1.6in]{goto}}
  \caption{POM of Probabilistic {\gt{Coincidence}} and {\gt{Exclusion}}}
\label{fig:pomsc}
\end{figure}

\vspace{0.1in}
\noindent\textbf{Probabilistic Subclock} is employed to specify \ed\ {\gt{Periodic}} timing constraint, given as $signRecTrig\ \textcolor{red}{\subseteq_{p}}\ cTrig$ ({{Spec. R2}} in Fig.\ref{fig:East-adl model}). The corresponding POM is shown in Fig.\ref{fig:pomsub}:
The \staf\ chart {${Obs}$} in Fig.\ref{fig:pomsub}.(b) is utilized for \emph{Hypothesis Testing}, where $k$ is the total number of ticks of $signRecTrig$ ({{subclock}}) and $m$ is the number of ticks satisfying the {\gt{subclock}} \emph{relation}.
\noindent Whenever $signRecTrig$ ticks, $k$ is increased by 1, and if the {\gt{subclock}} \emph{relation} holds on that tick (i.e., the condition ``$signRecTrig \Longrightarrow cTrig$' is true), $m$ is increased by 1. When $k$ is increased to the sample size $N$, the \staf\ chart then judges whether the number of ``success'' ticks of $signRecTrig$ is greater than or equal to ``$p*k$'' (i.e., whether $\frac{m}{k} \geq p$ is valid), and it activates either \emph{valid} (``H$_0$'' is accepted) or \emph{fail} state (``H$_1$'' is accepted). A {\scriptsize $\ll$}Proof Objective{\scriptsize $\gg$} block with \emph{false} value is employed to check whether the probabilistic {\gt{subclock}} \emph{relation} is satisfied, i.e., the \emph{fail} is never reached. In similar, using the {${Obs}$} chart, other Pr\ccsl\ \emph{relations} can be represented as POMs. Further details are given below.
\vspace{0.1in}
\noindent\textbf{Probabilistic Coincidence} is adapted to specify \ed\ {\gt{Periodic}} timing constraint, given as $cTrig\ \textcolor{red}{\equiv_{p}}\ \{{\gt{PeriodicOn}}\ ms\ {\gt{period}}\ 50\}$ ({{Spec. R1}} in Fig.\ref{fig:East-adl model}). The representative POM is shown in
Fig.\ref{fig:pomsc}.(a):
A {\gt{PeriodicOn}} subsystem (whose internal blocks are shown in Fig.\ref{fig:periodic}) is utilized to generates a periodic clock $res$ that ticks every 50ms.
According to Definition 4 in Sec. 4, if either $cTrig$ or $res$ ticks
 (``$cTrig\ {\gt{OR}}\ {res}$'' is true),
$c$ becomes true and $k$  is increased by 1.
Meanwhile, if $cTrig$ and $res$ tick simultaneously (``${cTrig}\ {\gt{AND}}\ {res}$'' is true), $r$ becomes true and $m$ is increased by 1.
Based on the value of $m$ and $k$, {${Obs}$} checks whether the probability of {\gt{coincidence}} \emph{relation} being satisfied is greater than or equal to \emph{p} and
activates either ${valid}$ or ${fail}$ state. {\scriptsize $\ll$}Proof Objective{\scriptsize $\gg$} block checks whether ${fail}$ state is always inactive, i.e., H$_0$ is accepted.

\vspace{0.1in}
\noindent\textbf{Probabilistic Exclusion} is utilized to specify \ed\ {\gt{Exclusion}} timing constraint, given as $turnLeft\ \textcolor{red}{\#_p}\ rightOn$ (Spec. R27 in Fig.\ref{fig:East-adl model}). The corresponding POM  is shown in Fig.\ref{fig:pomsc}.(b): $k$ is increased by 1 when
either $turnLeft$ or $rightOn$ ticks. If only one of the two clocks ticks at the current step, i.e., $r$ (the input of $Obs$) is true,
$m$ is increased by 1.
{\scriptsize $\ll$}Proof Objective{\scriptsize $\gg$} block with false value checks whether $fail$ state is never reached, i.e., H$_0$ is accepted.

\begin{figure}[htbp]
  \centering
  \includegraphics[width=5in]{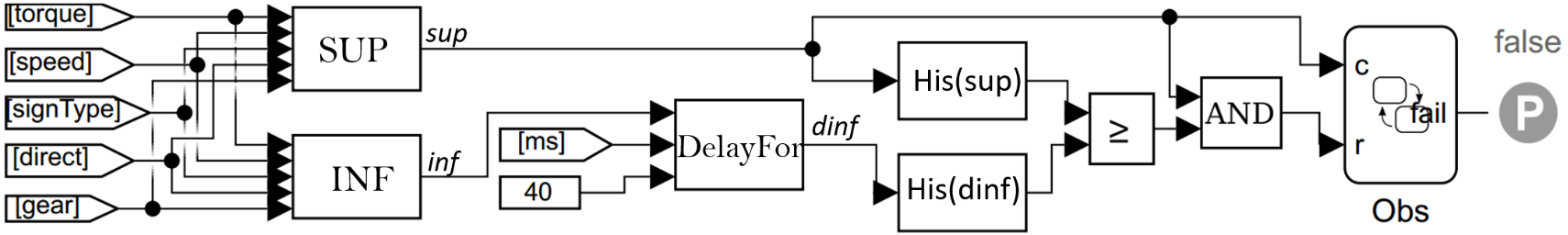}\\
  \caption{$sup\ \textcolor{red}{\preceq_{ p}}\ \{inf\ {\gt{DelayFor}}\ 40\ {\gt{on}}\ ms\}$}
  \label{fig:cause}
\end{figure}

\vspace{0.1in}
\noindent\textbf{Probabilistic Causality} is employed to specify \ed\ {\gt{Synchronization}} timing constraint,
$sup\ \textcolor{red}{\preceq_{ p}}\ \{inf\ {\gt{DelayFor}}\ 40\ {\gt{on}}\ ms\}$ (Spec. R13 in Fig.\ref{fig:East-adl model})),  where \emph{sup} (\emph{inf}) is the fastest (slowest) event slower (faster) than the five input events,  \emph{speed}, \emph{signType}, \emph{direct}, \emph{gear} and \emph{torque}. $sup$ and $inf$ are defined as:
\begin{equation}
sup\ \triangleq\ {{\gt{Sup}}}({{\gt{Sup}}}(speed,\ signType),\ {{\gt{Sup}}}({\gt{Sup}}(direct,\ gear),\ toque))
\end{equation}
\begin{equation}
inf\ \triangleq\ {\gt{Inf}}({\gt{Inf}}(speed,\ signType),\ {\gt{Inf}}({\gt{Inf}}(direct,\ gear),\ toque))
\end{equation}
The representative POM is illustrated in Fig.\ref{fig:cause}: The S/S models of {\gt{Inf}} and {\gt{Sup}} (shown in Fig.\ref{fig:infsup}) are utilized in order to construct
$inf$ (54) and $sup$ (55), modeled as {\gt{INF}} and {\gt{SUP}} subsystems, respectively.
A new clock \emph{dinf} is generated by delaying \emph{inf} for 40 ticks of $ms$, i.e.,
$dinf \triangleq \{inf\ {{\gt{DelayFor}}}\ 40\ {\gt{on}}\ ms\}$, and it is represented by using the S/S model of {{\gt{DelayFor}}} (shown in Fig.\ref{fig:delay}). Then {\gt{Probabilistic  Causality}} \emph{relation} between $sup$  and $dinf$ is checked.
According to Definition 6, when $sup$ ticks, $k$ is increased by 1.
At the same step, if the {\gt{causality}} \emph{relation} between $sup$ and $dinf$ is satisfied, i.e., the history of $sup$ is greater than
or equal to the  history of $dinf$, $m$ is increased by 1.
{\scriptsize $\ll$}Proof Objective{\scriptsize $\gg$} block analyzes if the {\gt{Probabilistic Causality}} \emph{relation} is satisfied , i.e., the $fail$ state is never activated.

\begin{figure}[htbp]
\centering
  \subfigure[ \{$imIn$ \gt{DelayFor} 100 \gt{on} $ms$\} \ \textcolor{red}{$\preceq_{p}$} $signOut$]{
  \includegraphics[width=3.7in]{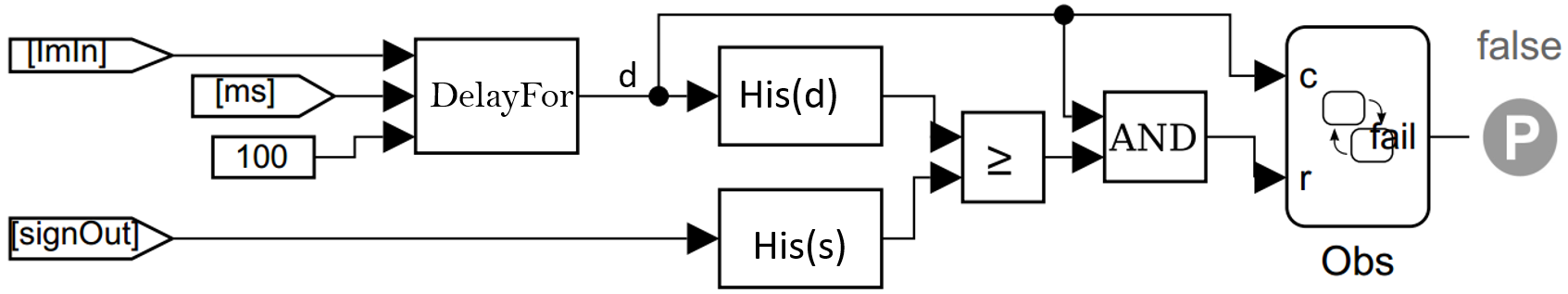}}
  \subfigure[$signOut$ \textcolor{red}{$\preceq_{0.95}$} \{$imIn$ \gt{DelayFor} 150 \gt{on} $ms$\}]{
  \includegraphics[width=3.7in]{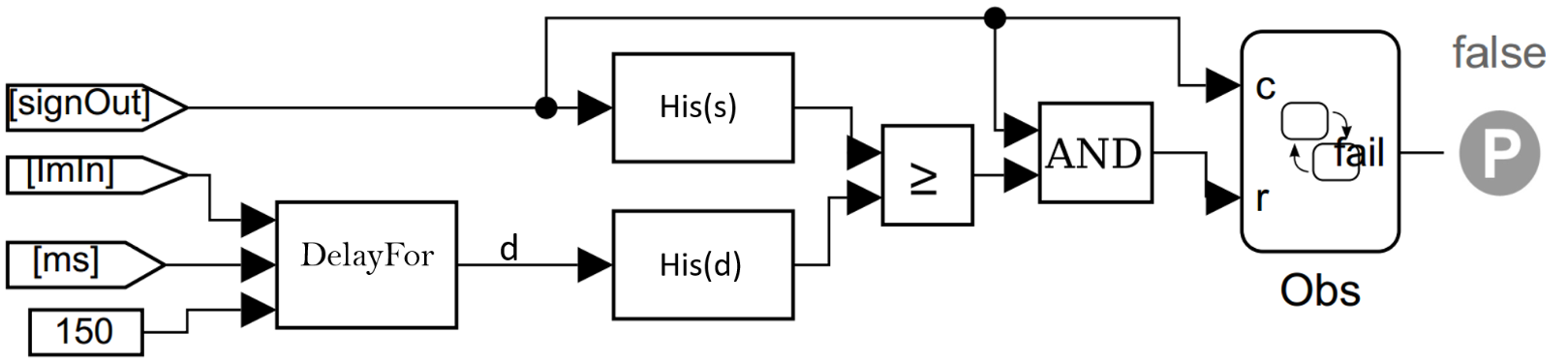}}
 % \subfigure[From and Goto]{
  %\includegraphics[width=1.6in]{goto}}
  \caption{POM of Probabilistic {\gt{Coincidence}} and {\gt{Exclusion}}}
\label{fig:pomexe}
\end{figure}

\noindent In Similar, \ed\ {\gt{Execution}} (R5) can be specified in {\gt{Probabilistic Causality}} using {\gt{DelayFor}} and translated into corresponding POMs. The \gt{execution} timing constraint R5 can be divided into two sub-properties, given as R5(1) and R5(2) in Sec.\ref{sec: specification}.
The POM models of R5(1) and R5(2) are illustrated in Fig.\ref{fig:pomexe}.(a) and Fig.\ref{fig:pomexe}.(b) respectively. Two intermediate clocks are generated by delaying $imIn$ for 100 ticks and 150 ticks of $ms$ (the output of the \emph{DelayFor} subsystem).
Then the \gt{execution} timing constraints, interpreted as the {\gt{probabilistic causality}} \emph{relation}, can be modeled with \emph{Obs} chart.

\begin{figure}[htbp]
  \centering
  \includegraphics[width=5.2in]{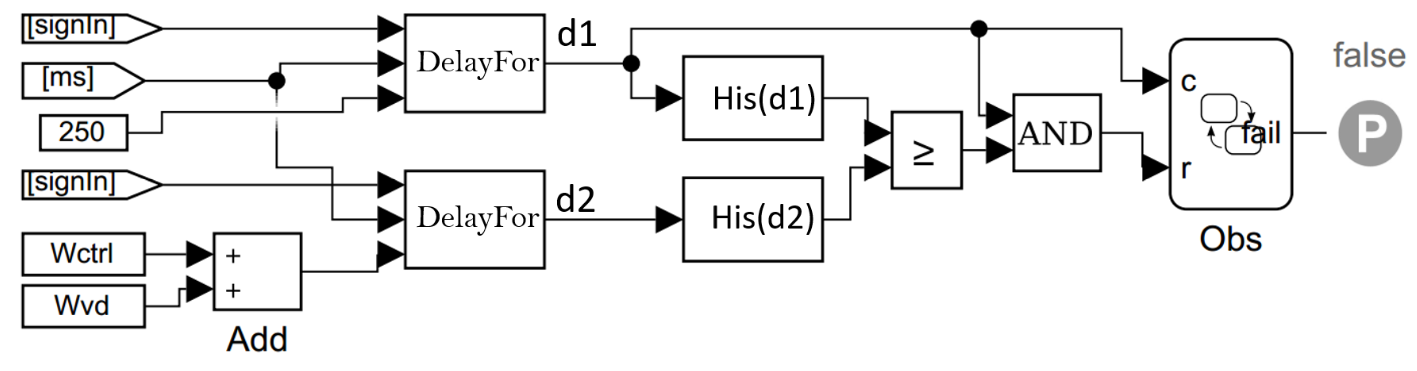}\\
  \caption{\{$signIn$ \gt{DelayFor} 250 \gt{on} $ms$\} \textcolor{red}{$\preceq_{p}$} \{$signIn$ \gt{DelayFor} ($Wctrl$ + $Wvd$) \gt{on} $ms$\}}
  \label{fig:compom}
\end{figure}

\noindent {\gt{Comparison}} (R24) timing constraint, specified in {\gt{Probabilistic Causality}} and {\gt{DelayFor}} (see Sec. \ref{sec: specification}), can be translated into the POMs presented in Fig.\ref{fig:compom}. Two intermediate clocks are generated by using the S/S model of {\gt{DelayFor}}, i.e., $d1$ is the $signIn$ clock delayed for 250 and $d2$ is the clock generated by delaying $signIn$ for ($Wctrl$ + $Wvd$) ticks of $ms$. Afterwards, the $Obs$ is applied to check whether the {\gt{Probabilistic Causality}} \emph{relation} between $d1$ and $d2$ is satisfied, i.e., whether the history of $d1$ is always greater than or equal to the history of $d2$.

\begin{figure}[htbp]
  \centering
  \includegraphics[width=4.8in]{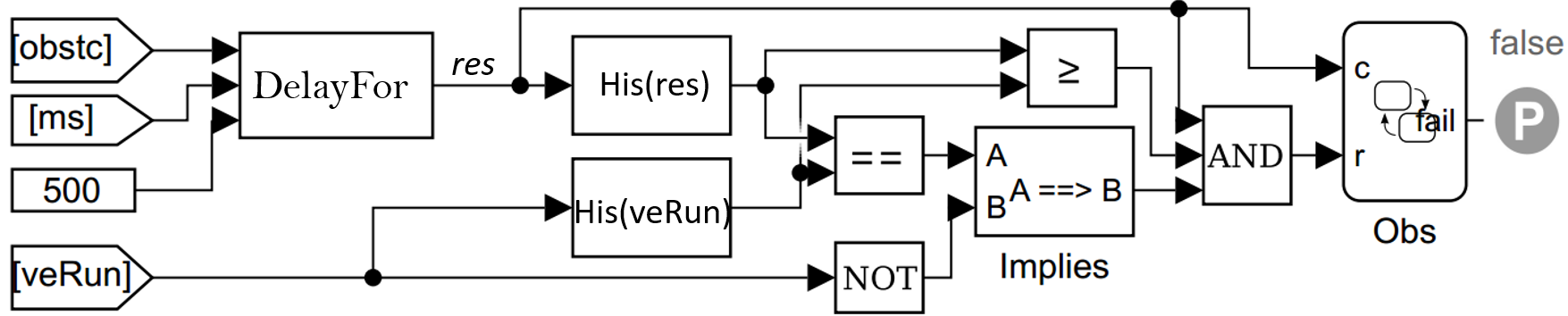}\\
  \caption{$\{obstc\ {\gt{DelayFor}}\ 500\ {\gt{on}}\ ms\}\ \textcolor{red}{\prec_{ p}}\ veRun$ }
  \label{fig:prec}
\end{figure}

\vspace{0.1in}
\noindent\textbf{Probabilistic Precedence}  is used to specify \ed\ {\gt{Sporadic}} timing
constraint, given as $\{obstc\ {\gt{DelayFor}}\ 500\ {\gt{on}}\ ms\}\ \textcolor{red}{\prec_{ p}}\ veRun$ (Spec. R9 in Fig.\ref{fig:East-adl model}).
The constraint delimits that two events $obstc$ and $veRun$ must have a minimum delay 500ms, and its corresponding POM is illustrated in Fig.\ref{fig:prec}: A new clock $res$ is generated by delaying $obstc$ by 500 ticks of $ms$, i.e.,
$res$ $\triangleq$ $\{obstc$ {{\gt{DelayFor}}} $500$ {\gt{on}} $ms\}$, and it is modeled by using the S/S model of {\gt{DelayFor}}. Then R9 can be checked by verifying $res\ \textcolor{red}{\prec_{ p}}\ veRun$. As presented in Fig.\ref{fig:prec}, whenever $res$ ticks, $c$ becomes true and $k$ is increased by 1. If the tick of $obstc$ satisfies the \gt{precedence} \emph{relation}, i.e., the history of $res$ is greater than or equal to the history of $veRun$ (excludes $res$ and $veRun$ are coincident), $r$ becomes true and $m$ is be increased by 1. {\scriptsize $\ll$}Proof Objective{\scriptsize $\gg$} block checks whether {\gt{Probabilistic Precedence}} is satisfied, i.e., the $fail$ state is never activated.

\begin{figure}[htbp]
\centering
  \subfigure[ \{$signIn$ \gt{DelayFor} 150 \gt{on} $ms$\} \textcolor{red}{$\prec_{p}$} $tqOut$ ]{
  \includegraphics[width=4.2in]{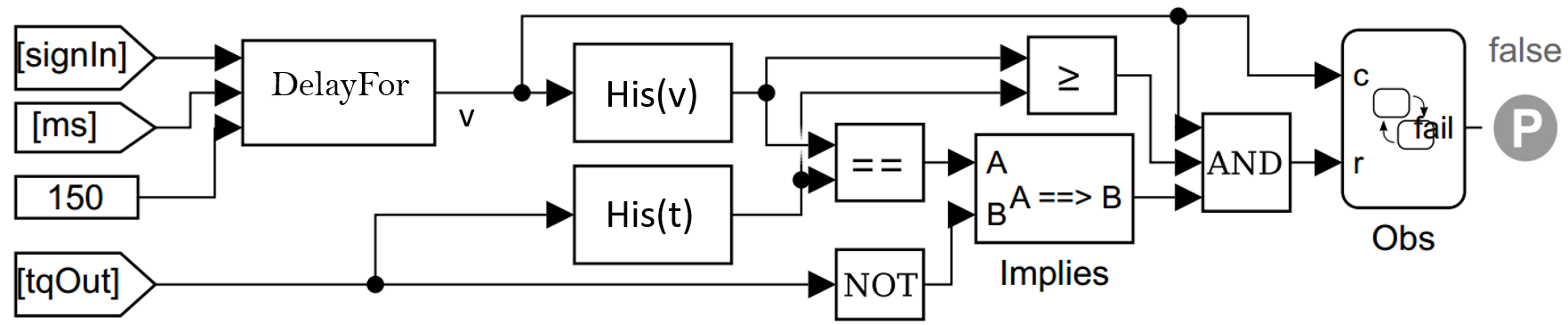}}
  \subfigure[ $tqOut$ \textcolor{red}{$\prec_{p}$} \{$signIn$ \gt{DelayFor} 250 \gt{on} $ms$\}]{
  \includegraphics[width=4.2in]{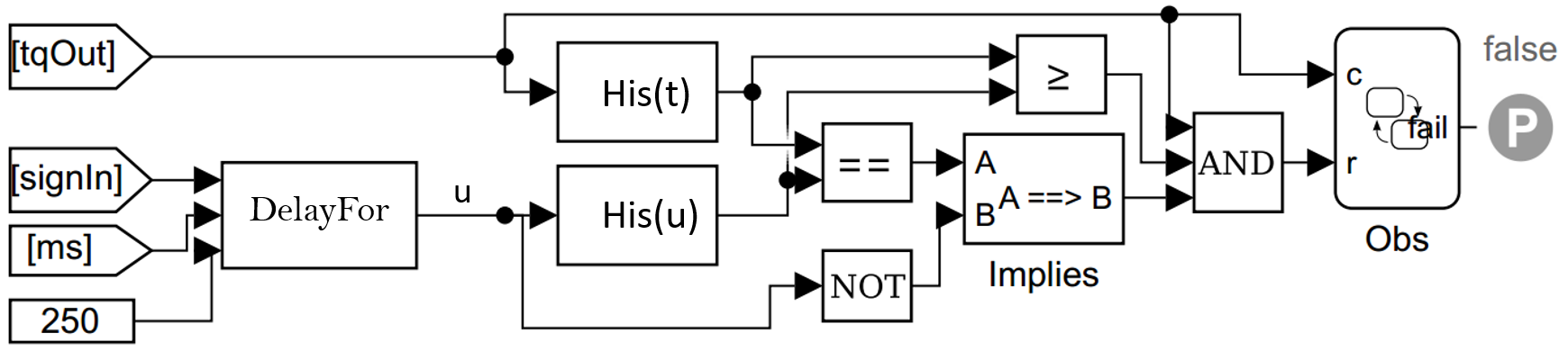}}
 % \subfigure[From and Goto]{
  %\includegraphics[width=1.6in]{goto}}
  \caption{POM of {\gt{End-to-End}} timing constraint }
\label{fig:pome2e}
\end{figure}

\noindent Similarly, {\gt{End-to-End}} timing constraint (R17) specified in {\gt{Probabilistic
}}\gt{{Prec}}-\gt{{edence}} (see Sec.\ref{sec: specification}) can be translated into its corresponding POM.
The constraint R17 can be divided into two sub-properties, R17(1) and R17(2) (see Sec.\ref{sec: specification}).
The corresponding POM of R17(1) and R17(2) are presented in Fig.\ref{fig:pome2e}.
For R17(1), a new clock $v$ (the output of \emph{DelayFor} subsystem) is generated by using the S/S model of {\gt{DelayFor}} such that the ticks of $v$ is the ticks of $signIn$ delayed for $150$ ticks of $ms$. To check whether R17(1) is satisfied is to verify whether $v$ always precedes $signOut$.
For R17(2), a new clock $u$ is constructed by delaying $signIn$ for 250 ticks on $ms$. The $Obs$ chart is then utilized to check whether the {\gt{Probabilistic
Precedence}} between $signOut$ and $u$ is satisfied.

\chapter{Modeling of AV System and its Environment in S/S}
\label{sec: av_ss}
\begin{figure}[htbp]
  \centering
  \includegraphics[width=4.7in]{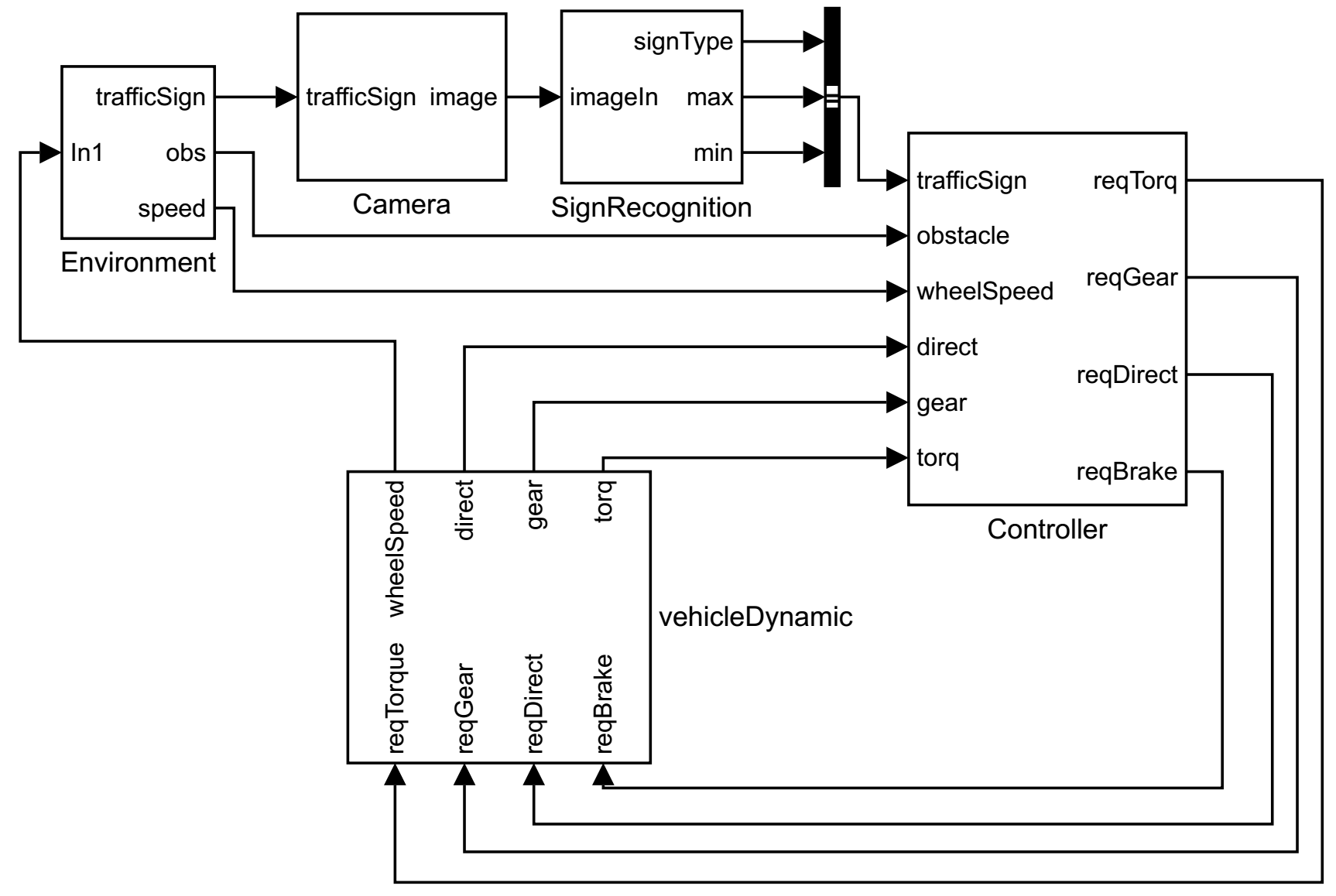}\\
  \caption{Top-view of AV in S/S}
  \label{fig:ssmodel}
\end{figure}

We have presented how the \ed\ timing constraints, specified in Pr\ccsl\ \emph{relations} and \ccsl\ \emph{expressions} are converted to POMs. To enable verification of the timed and stochastic behaviors of AV using \sdv, the behaviors of each \fp\ is described in S/S.   The {\gt{FA}}$_{SYS}$, consisting of a set of S/S is considered the entire behavior model of AV. The top-view architecture of  {\gt{FA}}$_{SYS}$ in S/S is shown in Fig.\ref{fig:ssmodel}.
\begin{figure}[htbp]
\centering
  \subfigure[\gt{Camera}]{
  \includegraphics[width=3.3in]{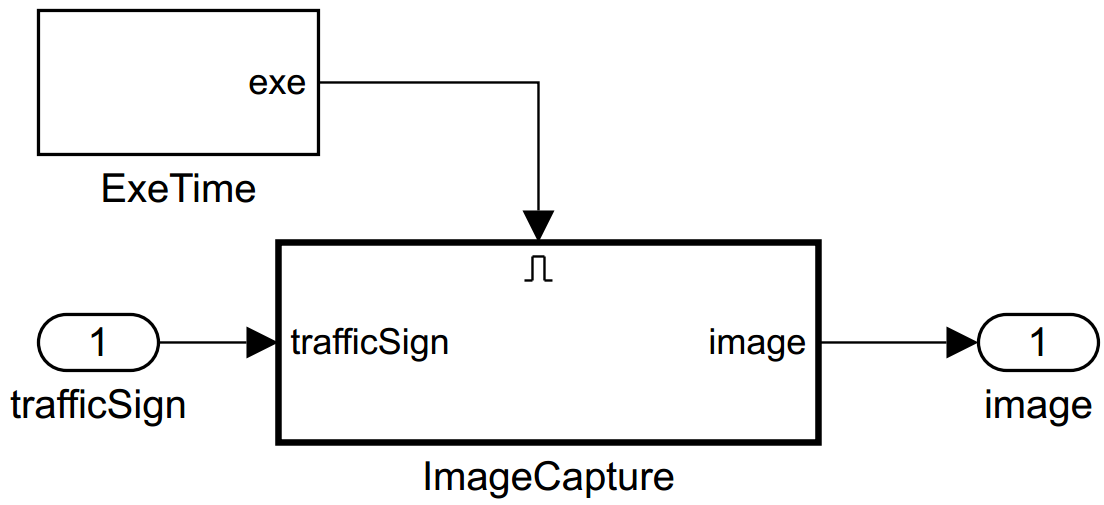}}
  \subfigure[\gt{SignRecognition}]{
  \includegraphics[width=3.3in]{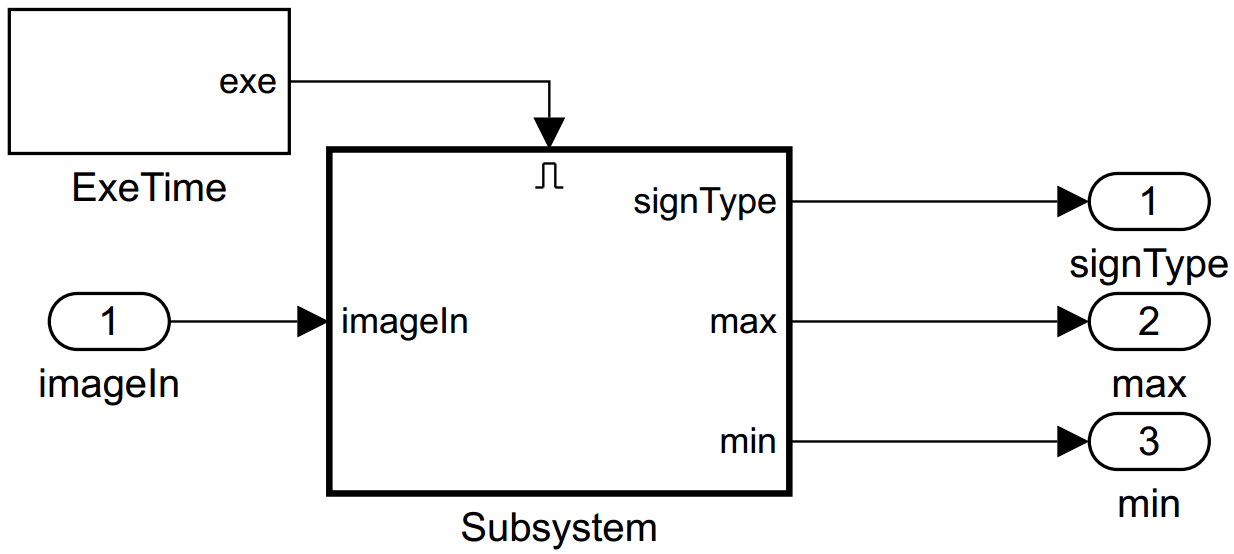}}
  \caption{Simulink model of \gt{Camera} and \gt{SignRecognition}}
\label{fig:cmrsr}
\end{figure}
\begin{figure}[htbp]
\centering
  \subfigure[Top-view of Stateflow chart]{
  \includegraphics[width=4in]{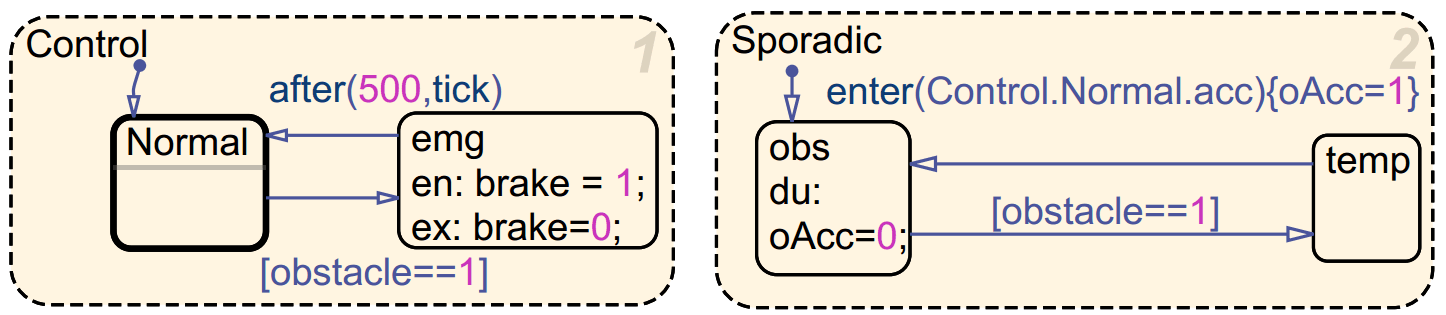}}
  \subfigure[Internal behaviors of Normal state]{
  \includegraphics[width=5in]{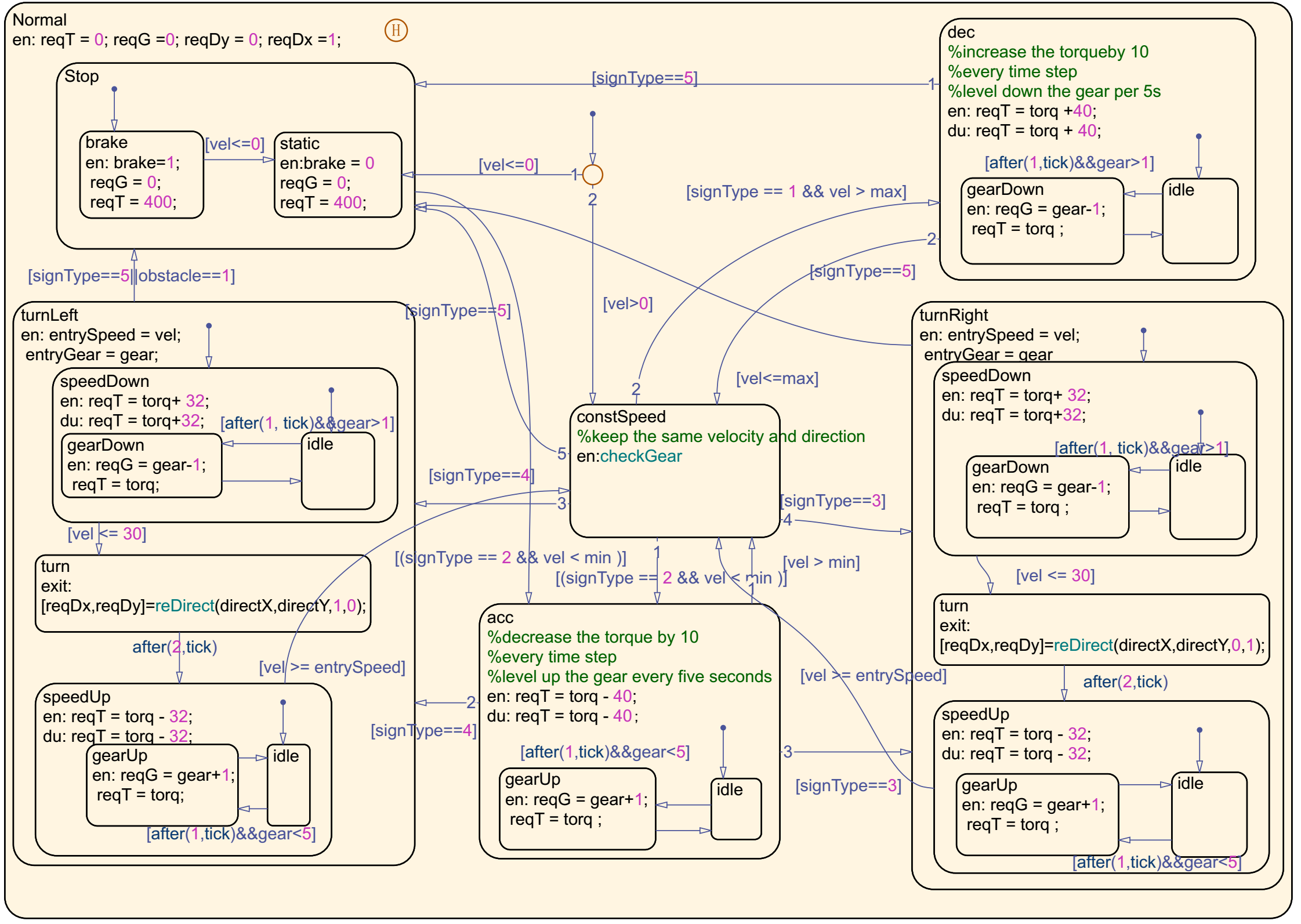}}
  \caption{Stateflow chart of  \gt{Controller} }
\label{fig:stateflow}
\end{figure}

Each \fp\ in \ed\ model is modeled in a {\scriptsize $\ll$}Subsystem{\scriptsize $\gg$} with input and output ports for communication with other \fp s.
To describe the stochastic environments of AV (modeled in the \emph{Environment} subsystem in Fig.\ref{fig:ssmodel}), a pseudo random number generator, \emph{Mersenne Twister} \cite{Matsumoto1998Mersenne} implemented in MATLAB script is employed: \begin{inparaenum} \item The traffic signs (6 types) are randomly recognized by AV and the probability of each sign type occurred is equally set as 16.7\%; \item  The probability of AV being obstructed by any obstacles is set to maximum 5\%; \item Since AV runs under different road conditions, speed variation influenced by the conditions ranges within [0, 2] m/s. \end{inparaenum}

The S/S model of {\gt{Camera}} and {\gt{SignRecognition}} are illustrated in Fig.\ref{fig:cmrsr}.(a) and Fig.\ref{fig:cmrsr}.(b) respectively.
Since {\gt{Camera}} and  {\gt{SignRecognition}} are triggered to execute periodically, \emph{ExeTime} subsystem is utilized to generate a boolean signal that becomes true periodically that can be the trigger signal of the {\gt{Camera}} and {\gt{SignRecognition}} subsystem. In {\gt{SignRecognition}}, the computation of traffic sign type of the detected image is implemented in a {\scriptsize $\ll$}Matlab Function{\scriptsize $\gg$} block.

As shown in Fig.\ref{fig:stateflow}, a Stateflow chart is employed to model the control logic of {\gt{Controller}}.   Fig.\ref{fig:stateflow}.(a) presents the top-view of
the {\gt{Controller}}, which consists of two parallel states (in ``AND'' decomposition), \emph{Control} and \emph{Sporadic}. If the vehicle is in the emergency mode because of encounter of obstacles, \emph{emg} state will be activated. Otherwise the \emph{Normal} state will be activated. There are five substates inside \emph{Normal} states (see Fig.\ref{fig:stateflow}.(b)), i.e., \emph{turnLeft} (the vehicle is turning left), \emph{turnRight} (the vehicle is turning right), \emph{Stop} (the vehicle is braking to stop), \emph{dec} (the vehicle is decelerating) and \emph{acc} (the vehicle is accelerating).

\begin{figure}[htbp]
\centering
  \subfigure[Top view of \gt{VehicleDynamic}]{
  \includegraphics[width=3in]{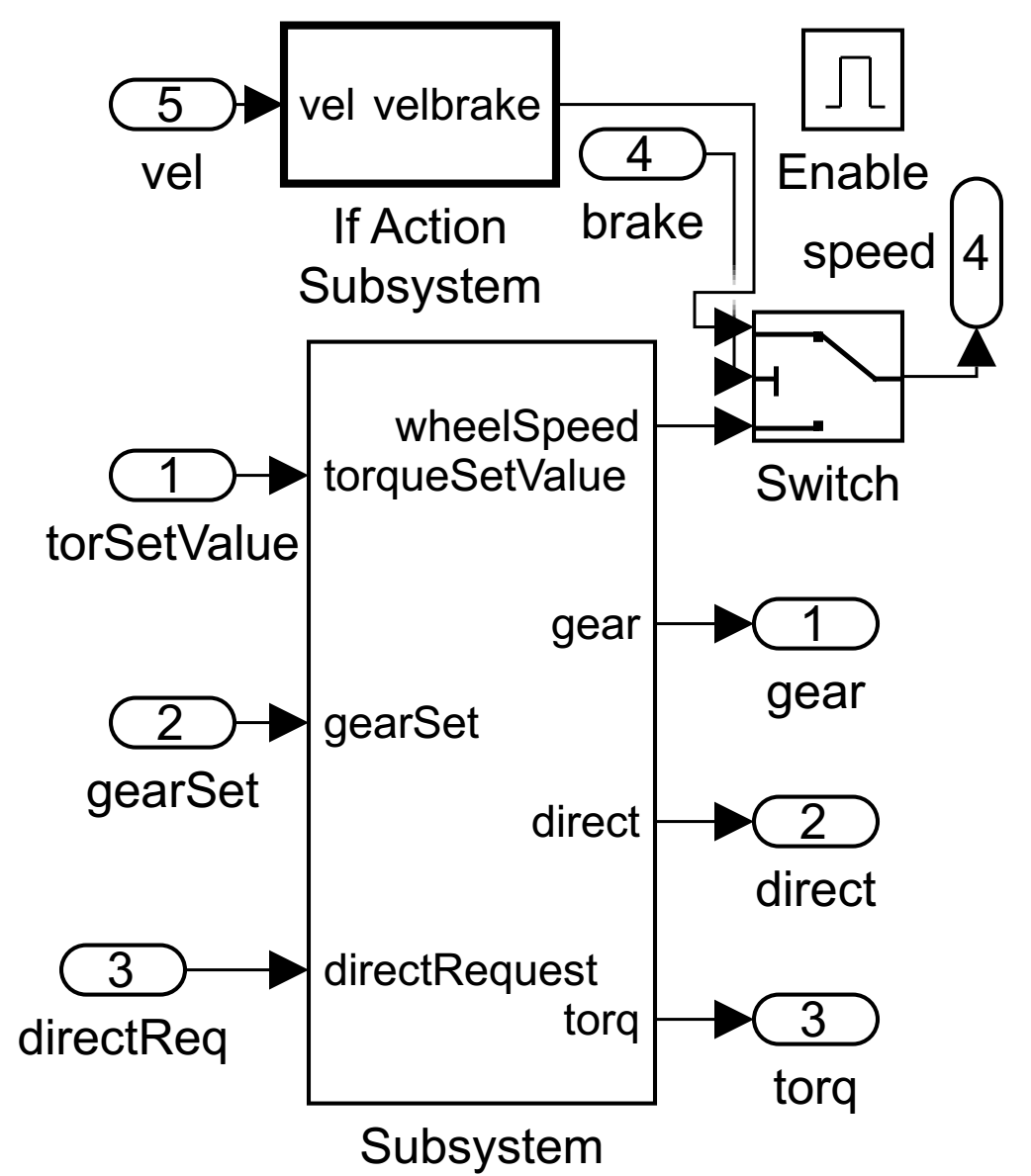}}
  \subfigure[Internal behaviors of \emph{Subsystem}]{
  \includegraphics[width=3in]{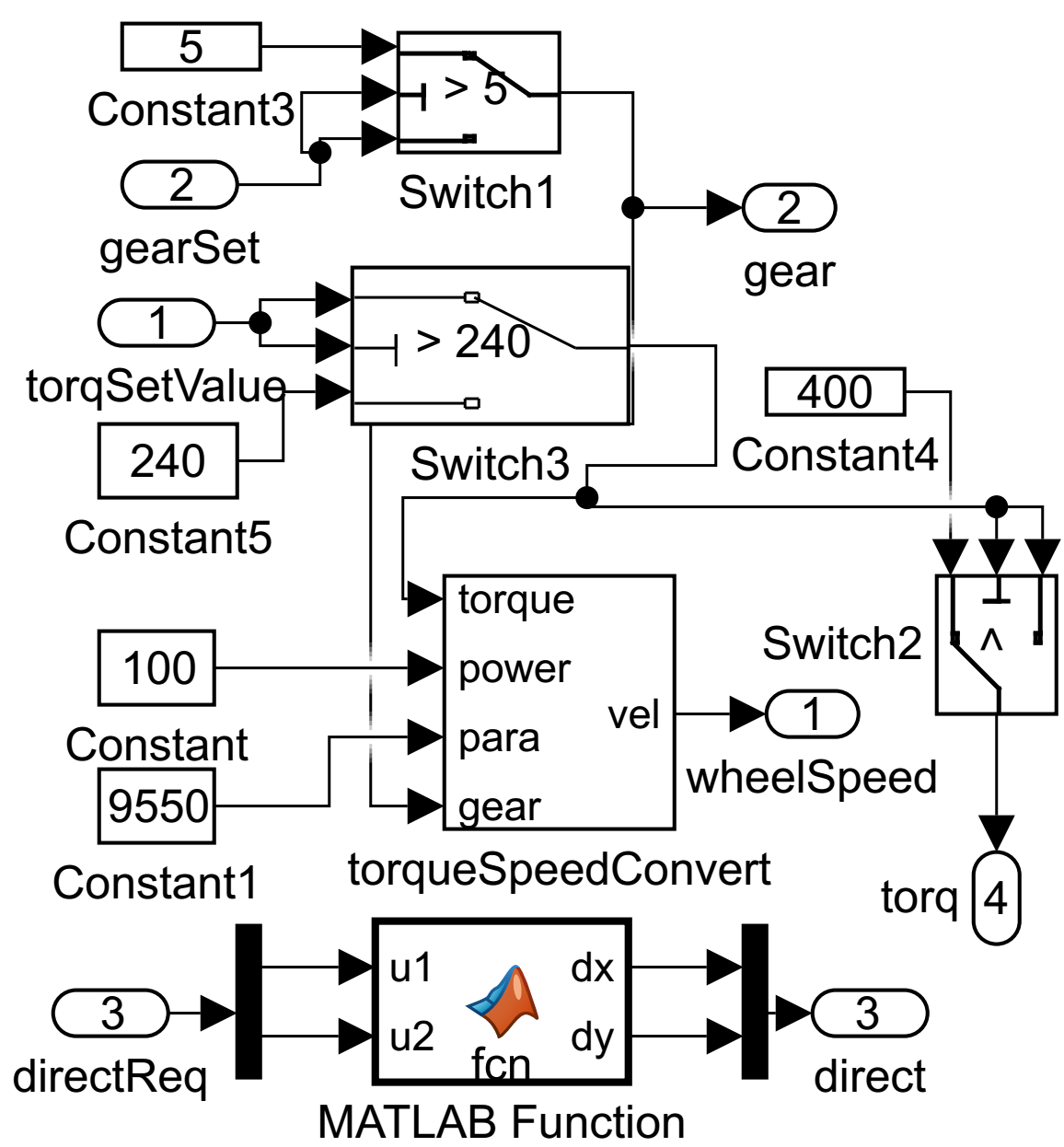}}
  \caption{Simulink model of \gt{VehicleDynamic} \fp}
\label{fig:vdfp}
\end{figure}

The inner behaviors of {\gt{VehicleDynamic}} in S/S is illustrated in Fig.\ref{fig:vdfp}.
{\gt{Vehic-}}  {\gt{leDynamic}} updates the speed and running direction of the vehicle according to the requests/commands of torque, gear and direction from {\gt{Controller}}.

\chapter{Experiments: Verification \& Validation}
\label{sec:verification}
We have formally specified and analyzed over 30 properties (associated with timing constraints) of the AV system. The properties (given in Sec. 3) are verified using \sdv\ and the results are listed in Table.\ref{table_verification_result}. The simulation bound and the probability threshold are set to 60000 steps and 95\% respectively. Maximum 4 properties per each \ed\ timing constraint are verified and all properties are established as valid. For further details regarding the full POMs and S/S models used in the experiment, refer to  \cite{prccsllibrary}.

\begin{table*}[htbp]
 \scriptsize
  \centering
   \renewcommand\arraystretch{1.3}
  \caption{Consolidated Verification Results in SDV}
    \begin{tabular}{|c|c|p{170pt}|c|p{23pt}|p{23pt}|p{23pt}|}
    \hline
    Category  & R.ID & Expression & Result & Time (Min) & Mem (Mb) & CPU (\%) \\
    \hline
        \multirow{4}{*}{Periodic} & R1 &$cTrig$ \bm{$\equiv_{0.95}$} \{\textbf{PeriodicOn} $ms$ \textbf{period} 50\} & valid & 6.28 & 2491 & 24.7 \\ \cline{2-7}
          &R2 & $signTrig$ \bm{$\equiv_{0.95}$} \{\textbf{PeriodicOn} $ms$ \textbf{period} 200\} & valid & 6.36 & 3920 & 24.13\\ \cline{2-7}
          & R3& $obsDetect$ \bm{$\equiv_{0.95}$} \{\textbf{PeriodicOn} $ms$ \textbf{period} 40\} & valid & 6.35 & 2357 & 24.7\\ \cline{2-7}
          & R4 &$spUpdate$ \bm{$\equiv_{0.95}$} \{\textbf{PeriodicOn} $ms$ \textbf{period} 30\}  & valid & 7 & 2218 & 24.01
    \\ \hline
    \multirow{4}{*}{Execution} & R5 &\{$imIn$ \textbf{DelayFor} 100 \textbf{on} $ms$\} \ \bm{$\preceq_{0.95}$} $signOut$ & valid & 38.20 & 4086 & 24.73 \\ \cline{3-7}
          & & $signOut$ \bm{$\preceq_{0.95}$} \{$imIn$ \textbf{DelayFor} 150 \textbf{on} $ms$\} & valid & 33.96 & 16225 & 19.90\\ \cline{2-7}
          &R6 &\{$cmrTrig$ \textbf{DelayFor} 20 \textbf{on} $ms$\} \ \bm{$\preceq_{0.95}$} $cmrOut$ & valid & 44:26 & 14379 & 18.39\\ \cline{3-7}
          & & $cmrOut$ \bm{$\preceq_{0.95}$} \{$cmrTrig$ \textbf{DelayFor} 30 \textbf{on} $ms$\} & valid & 51.15 & 4428.6 & 24.89\\ \cline{2-7}
          &R7 & \{$ctrlIn$ \textbf{DelayFor} 100 \textbf{on} $ms$\} \ \bm{$\preceq_{0.95}$} $ctrlOut$ & valid & 62.83 & 18306 & 6.09\\ \cline{3-7}
          & & $ctrlOut$ \bm{$\preceq_{0.95}$} \{$ctrlIn$ \textbf{DelayFor} 150 \textbf{on} $ms$\} & valid & 63.88 & 10737 & 22.04\\ \cline{2-7}
          &R8 & -\{$vdIn$ \textbf{DelayFor} 50 \textbf{on} $ms$\} \ \bm{$\preceq_{0.95}$} $vdOut$& valid & 49.13 & 17705 & 6.40\\ \cline{3-7}
          & & $vdOut$ \bm{$\preceq_{0.95}$} \{$vdIn$ \textbf{DelayFor} 100 \textbf{on} $ms$\} & valid & 34.05 & 18511 & 6.02
    \\ \hline
            \multirow{4}{*}{Sporadic} & R9 &\{$obstc$ \textbf{DelayFor} 500 \textbf{on} $ms$\} \bm{$\prec_{0.95}$} $veRun$ & valid & 100.5 & 13961 & 18.05 \\ \cline{2-7}
          &R10 & \{$obstc$ \textbf{DelayFor} 500 \textbf{on} $ms$\} \bm{$\prec_{0.95}$} $veAcc$ & valid & 120.45 & 13873 & 17.99\\ \cline{2-7}
          & R11& \{$obstc$ \textbf{DelayFor} 500 \textbf{on} $ms$\} \bm{$\prec_{0.95}$} $tLeft$ & valid &106.89  &13775 & 16.94\\ \cline{2-7}
          & R12 & \{$obstc$ \textbf{DelayFor} 500 \textbf{on} $ms$\} \bm{$\prec_{0.95}$} $tRight$& valid &143.26  &13775  &16.07
    \\ \hline
    \multirow{4}{*}{Synchronization} & R13 &${sup_{ctrlIn}}$ \bm{$\preceq_{0.95}$} \{${inf_{ctrlIn}}$ \textbf{DelayFor} 40 \textbf{on} $ms$\} & valid & 38.95 & 14135 & 16.85\\ \cline{2-7}
          &R14 & ${sup_{ctrlOut}}$ \bm{$\preceq_{0.95}$} \{${inf_{ctrlOut}}$ \textbf{DelayFor} 30 \textbf{on} $ms$\} & valid & 42.6 & 20616 &18.32 \\ \cline{2-7}
          & R15& ${sup_{vdIn}}$ \bm{$\preceq_{0.95}$} \{${inf_{vdIn}}$ \textbf{DelayFor} 40 \textbf{on} $ms$\} & valid & 66.78 &2196 &23.36 \\ \cline{2-7}
          & R16 & ${sup_{vdOut}}$ \bm{$\preceq_{0.95}$} \{${inf_{vdOut}}$ \textbf{DelayFor} 40 \textbf{on} $ms$\}& valid & 34.6 &3164  &24.07
    \\ \hline
        \multirow{4}{*}{End-to-End} & R17 &\{$signIn$ \textbf{DelayFor} 150 \textbf{on} $ms$\} \bm{$\prec_{0.95}$} $tqOut$  & valid & 35.95 & 6307 & 24.31 \\ \cline{3-7}
          & & $tqOut$ \bm{$\prec_{0.95}$} \{$signIn$ \textbf{DelayFor} 250 \textbf{on} $ms$\} & valid & 24.95 & 3989 & 24.07\\ \cline{2-7}
          &R18 & \{$cmrTrig$ \textbf{DelayFor} 120 \textbf{on} $ms$\} \bm{$\prec_{0.95}$} $signOut$  & valid & 33.96 & 6309 & 24.49\\ \cline{3-7}
          & & $signOut$ \bm{$\prec_{0.95}$} \{$cmrTrig$ \textbf{DelayFor} 180 \textbf{on} $ms$\} & valid & 43.02 & 6308 & 24.29\\ \cline{2-7}
          &R19 & \{$cmrTrig$ \textbf{DelayFor} 270 \textbf{on} $ms$\} \bm{$\prec_{0.95}$} $spOut$ & valid & 132.4 & 16287& 9.53\\ \cline{3-7}
          & & $spOut$ \bm{$\prec_{0.95}$} \{$cmrTrig$ \textbf{DelayFor} 430 \textbf{on} $ms$\} & valid & 163.8 & 16090 & 24.53\\ \cline{2-7}
          &R20 & $startTurnLeft$ \bm{$\prec_{0.95}$} \{$DetectLeftSign$ \textbf{DelayFor} 500 \textbf{on} $ms$\} & valid & 63.2 & 13052 & 12.74\\ \cline{2-7}
          &R21 & $startTurnRight$ \bm{$\prec_{0.95}$} \{$DetectRightSign$ \textbf{DelayFor} 500 \textbf{on} $ms$\} & valid & 76.5 & 15132 & 10.46\\ \cline{2-7}
          &R22 & $startBrake$ \bm{$\prec_{0.95}$} \{$DetectStopSign$ \textbf{DelayFor} 500 \textbf{on} $ms$\} & valid & 69 & 15293 & 9.38\\ \cline{2-7}
          &R23 & $Stop$ \bm{$\prec_{0.95}$} \{$DetectStopSign$ \textbf{DelayFor} 3000 \textbf{on} $ms$\} & valid & 95.7 & 15396 & 9.38
    \\ \hline
        \multirow{4}{*}{Comparison} & R24 &\{$signIn$ \textbf{DelayFor} 250 \textbf{on} $ms$\} \bm{$\preceq_{0.95}$} $\newline$ \{$signIn$ \textbf{DelayFor} ($Wctrl$ + $Wvd$) \textbf{on} $ms$\}\ & valid & 17.88 & 6309 & 24.61 \\ \cline{2-7}
          &R25 & \{$cmrTrig$ \textbf{DelayFor} 180 \textbf{on} $ms$\} \bm{$\preceq_{0.95}$} $\newline$ \{$cmrTrig$ \textbf{DelayFor} ($Wcmr$ + $Wsr$) \textbf{on} $ms$\}\ & valid & 60.15 & 6410 & 24.43\\ \cline{2-7}
          & R26 & \{$cmrTrig$ \textbf{DelayFor} 430 \textbf{on} $ms$\} \bm{$\preceq_{0.95}$} $\newline$ \{$cmrTrig$ \textbf{DelayFor} ($Wcmr$ + $Wsr$ + $Wctrl$ + $Wvd$) \textbf{on} $ms$\}\ & valid & 43.33 &17370  & 14.15
    \\ \hline
        \multirow{4}{*}{Exclusion} & R27 &$turnLeft$ \bm{$\#_{0.95}$} $rightOn$ & valid & 387.76 & 20987 & 8.25 \\ \cline{2-7}
          &R28 & $veAcc$ \bm{$\#_{0.95}$} $veBrake$ & valid & 360.15 & 21168 & 18.15\\ \cline{2-7}
          & R29& $emgcy$ \bm{$\#_{0.95}$} $turnLeft$ & valid & 233.6 & 22861 &11.98\\ \cline{2-7}
          & R30& $emgcy$ \bm{$\#_{0.95}$} $turnRight$ & valid & 498.51 & 23245 & 9.97\\ \cline{2-7}
          & R31 &$emgcy$ \bm{$\#_{0.95}$} $veAcc$ & valid & 260.96 & 22257 & 8.85
    \\ \hline
    \end{tabular}%
  \label{table_verification_result}%
\end{table*}%

\chapter{Related work}
\label{sec: related work}
Considerable research efforts have been devoted to formal analysis of CPS
by applying SDV \cite{gholami2016verifying, JFSLDV}, which are however, limited to the functional properties without consideration of non-functional properties, i.e., timing constraints.
In the context of \ed, efforts on the integration of \ed\ and formal techniques based on timing constraints were investigated in several works  \cite{kress13,qureshi2011,ksafecomp11,Goknil2013Analysis}, which are however, restricted to the executional aspects of system functions without addressing stochastic behaviors.
Kang \cite{ksac14} and Suryadevara \cite{Suryadevara2013Validating, Suryadevara2013Verifying} defined the execution semantics of both the controller and the environment of industrial systems in \ccsl\ which are given as mapping to \uppaal\ models amenable to model checking. In contrast to our current work, those approaches lack precise probabilistic annotations specifying stochastic properties.
Zhang \cite{Zhang2017Towards} transformed \ccsl\ into first order logics that are verifiable using SMT solver. However, this work is limited to functional properties, and no timing constraints are addressed.
Though, Kang et al. \cite{kiciea16,kapsec15} and Marinescu et al. \cite{Marinescu3762} presented both simulation and model checking approaches of \simu\ and \smc\ on \ed\ models, neither formal specification nor verification of extended \ed\ timing constraints with probability were conducted.
Our approach is a first application on the integration of \ed\ and formal V\&V techniques based on  probabilistic extension of \ed/\tdl\ constraints using \sdv.
An earlier study \cite{mvv, sscps, sac18}
defined a probabilistic extension of \ed\ timing constraints and presented model checking approaches on \ed\ models, which inspires our current work. Specifically, the techniques provided in this paper define new operators of \ccsl\ with stochastic extensions (Pr\ccsl) and formally verify the extended \ed\ timing constraints of CPS.
Du. et al. \cite{Du2016MARTE} proposed the use of \ccsl\ with probabilistic logical clocks to enable stochastic analysis of hybrid systems by limiting the possible solutions of clock ticks. Whereas, our work is based on the probabilistic extension of \ed\ timing constraints with the focus on probabilistic verification of the extended constraints, particularly, in the context of WH.
\chapter{Conclusion}
%In this paper, we present modeling techniques to model the Simulink/Stateflow model based on \ed~ by taking advantages of MetaEdit+ tool.  The translation patterns for the timing constraints translation in Simulink/Stateflow are presented. Moreover, we adopt the autonomous traffic sign recognition vehicle as the running example and investigate the methods for verification and validation of (non)-functional properties of the vehicle described in \ed~ model. Experiment of verification and validation against a wide set of requirements is conducted. We evaluate and compare verification result as well as the performance of SDV and \smc~tools. SDV only supports a finite set of Simulink blocks, which limits the capability for constructing the complex requirement when verification. However, \smc~can describe requirements for both controller and plant by using CTL expression. And it performs better than SDV in the aspect of time, memory and CPU consumption on average.
\label{sec: conclusion}

We present an approach to perform probabilistic analysis of \ed\ timing constraints in automotive systems at the early design phase: \begin{inparaenum} \item Probabilistic extension of \ccsl, called Pr\ccsl, is
defined and the \ed/\tdl\ timing constraints with stochastic properties
are specified in Pr\ccsl; \item The semantics of the extended constraints in Pr\ccsl, captured in \simu/\staf, is translated into verifiable POMs for formal verification; \item A set of mapping rules is proposed to facilitate guarantee of translation. \end{inparaenum} Our approach is demonstrated on an autonomous traffic sign recognition vehicle (AV) case study.
Although, we have shown that defining and translating a subset of \ccsl\ with probabilistic extension into POMs is sufficient to verify \ed\ timing constraints, as ongoing work, advanced techniques covering a full set of \ccsl\ constraints are further studied.
Despite the fact that SDV supports probabilistic analysis of the timing constraints of AV, the computational cost of verification in terms of time is rather expensive. Thus, we continuously investigate complexity-reducing design/mapping patterns for CPS to improve effectiveness and scalability of system design and verification.

\chapter*{Acknowledgment}
This work is supported by the National Natural Science Foundation of China and International Cooperation \& Exchange Program (46000-41030005) within the project EASY.

\end{document}